\documentclass[showpacs,prb,twocolumn]{revtex4}
\usepackage{amsfonts}
\usepackage{amssymb}
\usepackage{amsmath}
\usepackage{graphicx}
\begin{document}

\title{Nearest-Neighbor Repulsion and Competing Charge and Spin Order in the
Extended Hubbard Model.}
\author{B.~Davoudi$^{1,2}$ and A.-M.S.~Tremblay$^{1}$}

\begin{abstract}
We generalize the two-particle self-consistent approach (TPSC) to study the
extended Hubbard model, where nearest-neighbor interaction is present in
addition to the usual local screened interaction. Similarities and
differences between the TPSC approach and the Singwi Tosi Land Sj\"{o}lander
(STLS) approximation for the electron gas are discussed. The accuracy of our
extension of TPSC is assessed by comparisons with Quantum Monte Carlo
calculations of Y. Zhang and J. Callaway, Phys. Rev. B \textbf{39}, 9397
(1989). We observe that a positive off-site interaction enhances staggered
charge fluctuations and reduces staggered magnetic order.
\end{abstract}

\affiliation{$^{1}$D\'{e}partment de Physique and RQMP, Universit\'{e} de Sherbrooke,
Sherbrooke, Qu\'{e}bec, Canada J1K 2R1\\
$^{2}$Institute for Studies in Theoretical Physics and Mathematics, Tehran
19395-5531, Iran}

\maketitle

\section{Introduction}

\label{sec1}The electron gas with long-range Coulomb repulsion and the
Hubbard model with screened on-site repulsion are two widely studied models
that each describe large classes of materials. Both models are generally
studied with quite different theoretical methods. However, by gradually
increasing the range of the interaction in the Hubbard model and by reducing
the density to average out lattice effects, one should be able to go
continuously from the Hubbard to the Coulomb gas model. Is there one unified
theoretical framework that allows us to treat both limiting cases?

To begin to answer this question, we generalize the Two-Particle
Self-Consistent approach (TPSC) \cite{Vilk1, Vilk2} to study the extended
Hubbard model, that includes nearest-neighbor repulsion in addition to the
usual on-site repulsion. This model is interesting in its own right,
independently of the above-mentioned theoretical question. Indeed, the
extended Hubbard model allows one to study materials where competition
between charge and spin order manifest themselves. In high-temperature
superconductors, where screening is not perfect, understanding the extended
Hubbard model is also of paramount importance.

Let us first motivate further our focus on the TPSC approach and then come
back to the interesting physical phenomena that manifest themselves in the
extended Hubbard model. Judging from comparisons with benchmark Quantum
Monte Carlo calculations,\cite{Vilk1, Vilk2, Moukouri:2000, Kyung:2000,
Kyung:2003, Kyung:2003b} (in the absence of exact solutions), the TPSC
approach provides us with the most accurate approximate solution to the
Hubbard model at weak to intermediate coupling. A detailed critical
comparison with other methods such as the Random Phase Approximation (RPA),
the self-consistent renormalized theory and the fluctuation exchange
approximation is given in Ref.~(\onlinecite{Vilk2}). In particular, TPSC
satisfies the Pauli principle, the Mermin-Wagner theorem in two-dimensions
includes Kanamori-Br\"{u}ckner\cite{Kanamori-Bruckner} screening and is non-perturbative although
limited to interaction strengths less than the bandwidth. The TPSC approach
is a close relative to the Singwi, Tosi, Land, Sj\"{o}lander (STLS) method
\cite{Singwi} used in the electron gas problem. The STLS approximation has
been first introduced to describe the structure functions of an electron
liquid where it provided much better results compared to the RPA. This
approximation has been applied to a variety of systems that contain
Fermions, Bosons or classical particles in all physical dimensions and
different geometries. Starting with the equation of motion for one-body
density operator, the authors were faced with the well-known problem that
the two-body density operator appeared in their equation. They solved the
problem approximately by replacing the two-body density operator by a
product of two one-body density operators and then correcting the result
with the pair correlation function. The result of this factorization appears
as a correction in the response functions of the system, the so-called local
field factor. This factor is then determined by using a sum-rule derived
from the fluctuation-dissipation theorem. The present paper will give a new
point of view on the STLS method by comparing it to the TPSC approach. As we
will show in more details, the main difference between the latter method and
the STLS one, is the way we factorize the two-body density operator. It
seems that for local \textit{vs} non-local potentials, it is more accurate
to use, respectively, local or non-local factorization and, as we will show
in detail, a local factorization leads to better results for the extended
Hubbard model. That is not all. It will become clear in the formalism used
to derive TPSC that the STLS\ method also neglects some higher-order
correlation functions. The same type of approximation will be necessary to
be able to generalize TPSC to treat the extended Hubbard model\cite{fp}. Otherwise,
as in STLS, there is a shortage of sum-rules or conditions to find all
unknowns that appear in the method.

We believe that accuracy of the approximation is crucial for a real
understanding of physical properties and for meaningful comparisons to
experiments. Bad approximations that agree with experiment only lead us
astray. That is why we will benchmark our extension of the TPSC approach
against the highly accurate results that can obtained by Quantum Monte Carlo
(QMC) simulations. We will carefully analyze the approximations involved in
the method and discuss other possibilities for improvement.

Back to the extended Hubbard model. This model has a long history, so we can
only discuss a small sample of the relevant literature. At half-filling,
when the on-site interaction strength $U$ tends to infinity so that
super-exchange $4t^{2}/U$ vanishes, the effect of the nearest-neighbor
repulsion $V$ is to lead to an effective ferromagnetic interaction between
localized spins. This is the physics of the so-called direct-exchange
mechanism. The physics is quite different when $U$ and $4V$ are of the same
order and both in the weak to intermediate-coupling regime, namely less than
or of the order of the bandwidth ($W=8t$ in $d=2$). In that case, there is a
competition between staggered charge and spin orders. That
charge ordering phenomenon is particularly relevant for manganites,
vanadates and various organic conductors, as discussed in a recent
theoretical paper \cite{Imada:2005} that uses a new correlator-projection
method. The relevant theoretical literature for these compounds focuses on
the square lattice \cite{Onari:2004, Pietig:1999, Calandra:2002} and on
ladders for the quarter-filled case \cite{Vojta:1999, Vojta:2001}. The
competition between charge and spin orders has also been
studied in one- \cite{Fourcade, Bosch:1988, Aichhorn:2004} two- \cite%
{Avella:2004, Ohta:1994, Yan:1993}, three- \cite{Barktowiak:1995} and higher
dimensions \cite{VanDongen:1994} at various fillings. The combined effect of
charge fluctuations in addition to the usual spin fluctuation in favouring
one type or another of unconventional superconductivity has also been
studied using this model \cite{Onari, Kishine:1995, Callaway:1990}.

In the present paper, we are interested in the possibility that a
generalization of TPSC to the extended Hubbard model can lead us to accurate
estimations of the charge and spin structure factors and susceptibilities at
finite temperature. We will use the QMC calculations of Ref.~(\onlinecite{Zhang}) 
as a benchmark. We note that methods that have been quite
successful in one- or infinite dimension are generally not applicable in the
two-dimensional case that we will consider. In $d=2$, continuous symmetries
can be broken only at zero temperature and, in addition, wave-vector
dependencies that are neglected in high-dimension are generally not
negligible.

In the following, we first present the theory and give the details of the
calculation based on the functional derivative of the Dyson equation which
gives us the response functions of the system. We also provide the equation
of motion for the Wigner distribution function to show that the two
different methods basically lead to the same set of equations. This also
allows us to discuss the different types of factorizations. In Sec.~\ref{sec4}
we present the results of our numerical calculations and compare them
with QMC results to find the region where the method works properly or is
precise enough. Finally, we discuss the influence of nearest-neighbor
interaction $V$ on spin and charge fluctuations.

\section{Theory}

\label{sec2}We first introduce the extended Hubbard Hamiltonian,%
\begin{widetext}
\begin{equation}
H=-t\sum_{\left\langle \mathbf{ij}\right\rangle \sigma }(c_{\mathbf{i}\sigma
}^{\dagger }c_{\mathbf{j}\sigma }+c_{\mathbf{j}\sigma }^{\dagger }c_{\mathbf{%
i}\sigma })+U\sum_{\mathbf{i}}n_{\mathbf{i}\uparrow }n_{\mathbf{i}\downarrow
}+V\sum_{\left\langle \mathbf{ij}\right\rangle \sigma \sigma ^{\prime }}n_{%
\mathbf{i}\sigma }n_{\mathbf{j}\sigma ^{\prime }}-\mu \sum_{\mathbf{i}}n_{i}
\end{equation}%
\end{widetext}%
where $c_{\mathbf{i}\sigma }$ ($c_{\mathbf{i}\sigma }^{\dagger }$) are
annihilation (creation) operator for electrons of spin $\sigma $ at site $i$%
, $n_{\mathbf{i}\sigma }$ is the density operator, and $t$ is the hopping
matrix element. The quantities $U$ and $V$ are the on-site and
nearest-neighbor interactions respectively and $\mu $ is the chemical
potential. Although we restrict ourselves to nearest-neighbor hopping, the
generalization to an arbitrary hopping matrix $t_{i,j}$ will be obvious to
the reader. It only modifies the non-interacting dispersion relation. One
can generalize the formalism to a system with longer interaction terms and
also to a system with many bands. In the following, we first derive the TPSC
approach using functional derivatives \cite{Allen:2003} and then return to
the approach that is more usual with the STLS approximation \cite{Singwi},
namely the equation of motion for the one-body Wigner distribution function.
These two derivations allow a deeper insight into the nature of the
approximations. The reader may also choose the approach he is more familiar
with.


\subsection{Functional derivative approach}

Following functional methods of the Schwinger school \cite%
{Baym:1962,BaymKadanoff:1962,MartinSchwinger:1959}, we begin with the
generating function $\ln Z[\phi _{\sigma }]$ with source fields $\phi
_{\sigma }$ in the grand canonical ensemble%
\begin{equation}
Z[\phi _{\sigma }]=-\mathrm{Tr}\left[ e^{-\beta H}T_{\tau }S(\beta )\right]
\end{equation}%
where $\beta =1/T$, $T$ is the temperature
while $S$ is defined as follows
\begin{equation}
\ln S(\beta )=-\sum_{\mathbf{i},\mathbf{j},\sigma }\int_{0}^{\beta }d\tau
d\tau ^{\prime }c_{\mathbf{i}\sigma }^{\dagger }(\tau )c_{\mathbf{j}\sigma
}(\tau ^{\prime })\phi _{\sigma }(\mathbf{i},\mathbf{j},\tau ,\tau ^{\prime
})
\end{equation}%
where the brackets $\left\langle {}\right\rangle $ represents a thermal
average in the canonical ensemble, $T_{\tau }$ is the time-ordering
operator, and $\tau $ is the imaginary time. The Green function can be
calculated from the first functional derivative of the generating function $%
Z[\phi _{\sigma }]$ as follows%
\begin{equation}
G_{\sigma }(1,2)=-\frac{\delta \ln Z[\phi _{\sigma }]}{\delta \phi _{\sigma
}(2,1)}=\frac{\mathrm{Tr}\left[ e^{-\beta H}T_{\tau }S(\beta )c_{\sigma
}^{\dagger }(2)c_{\sigma }(1)\right] }{\mathrm{Tr}\left[ e^{-\beta H}T_{\tau
}S(\beta )\right] }  \label{GZ}
\end{equation}%
where we have introduced the short-hand $1$ to stand for both the site
position and the corresponding imaginary time, as in the equation%
\begin{equation}
G_{\sigma }(1,2)=-\left\langle T_{\tau }c_{\mathbf{1}\sigma }(\tau _{1})c_{%
\mathbf{2}\sigma }^{\dagger }(\tau _{2})\right\rangle =-\left\langle T_{\tau
}c_{\sigma }(1)c_{\sigma }^{\dagger }(2)\right\rangle .
\end{equation}

The equation of motion for the Green function has the following form \cite%
{BaymKadanoff:1962, Vilk2, Allen:2003},%
\begin{widetext}
\begin{equation}
G_{0}^{-1}(1,\bar{2})G_{\sigma }(\bar{2},3)=\delta (1,3)+\Sigma _{\sigma }(1,%
\bar{2})G_{\sigma }(\bar{2},3)+\phi _{\sigma }(1,\bar{2})G_{\sigma }(\bar{2}%
,3)  \label{emG}
\end{equation}%
where $G_{0}^{-1}(1,2)=[\frac{\partial }{\partial \tau }-\epsilon
(1,2)]\delta (1,2)$ is the non-interacting Green function, $\Sigma _{\sigma }
$ is the self-energy and the bar is a short-hand for $\sum_{\mathbf{i}}\int
d\tau $. The above equation is nothing more than the Dyson equation that can
be obtained by the diagrammatic technique. It can also be written in the
form
\begin{equation}
G_{\sigma }^{-1}(1,2)=G_{0}^{-1}(1,2)-\phi _{\sigma }(1,2)-\Sigma _{\sigma
}(1,2),  \label{Dyson}
\end{equation}%
with the self-energy
\begin{equation}
\Sigma _{\sigma }(1,2)=-U\left\langle T_{\tau }c_{\tilde{\sigma}}^{\dagger
}(1)c_{\tilde{\sigma}}(1)c_{\sigma }(1)c_{\sigma }^{\dagger }(\bar{3}%
)\right\rangle G_{\sigma }^{-1}(\bar{3},2)-V\sum_{\sigma ^{\prime
},a}\left\langle T_{\tau }c_{\sigma ^{\prime }}^{\dagger }(1+a)c_{\sigma
^{\prime }}(1+a)c_{\sigma }(1)c_{\sigma }^{\dagger }(\bar{3})\right\rangle
G_{\sigma }^{-1}(\bar{3},2)  \label{Self_exact}
\end{equation}%
where $\tilde{\sigma}=-\sigma $ and the summation on $a$ runs over the
nearest-neighbor sites of the site $1$ (the imaginary time is the same at $%
1+a$ as at $1$). 
One needs an approximation to deal with the above two-body density operator. By analogy
with the factorization pioneered by Singwi \textit{et. al}~~\cite{Singwi} we
write
\begin{align}
\Sigma _{\sigma }(1,2)& \cong UG_{\tilde{\sigma}}(1,1^{+})G_{\sigma }(1,\bar{%
3})G_{\sigma }^{-1}(\bar{3},2)g_{\sigma \tilde{\sigma}}(1,1)+V\sum_{\sigma
^{\prime },a}G_{\sigma ^{\prime }}(1+a,1+a^{+})G_{\sigma }(1,\bar{3}%
)G_{\sigma }^{-1}(\bar{3},2)g_{\sigma \sigma ^{\prime }}(1,1+a)  \notag \\
& =U\delta (1,2)G_{\tilde{\sigma}}(1,1^{+})g_{\sigma \tilde{\sigma}%
}(1,1)+V\delta (1,2)\sum_{\sigma ^{\prime },a}G_{\sigma ^{\prime
}}(1+a,1+a^{+})g_{\sigma \sigma ^{\prime }}(1,1+a)  \label{Self_approx}
\end{align}%
where $g_{\sigma \sigma ^{\prime }}(i,j)$ is the equal-time pair correlation
function which is related to the probability of finding one electron with
spin $\sigma ^{\prime }$ on site $j$ when another electron with spin $\sigma
$ is held on site $i$. More specifically,
\begin{equation}
g_{\sigma \sigma ^{\prime }}(1,2)\equiv \frac{\left\langle T_{\tau
}c_{\sigma }^{\dagger }(1)c_{\sigma }(1)c_{\sigma ^{\prime }}^{\dagger
}(2)c_{\sigma ^{\prime }}(2)\right\rangle -\delta (1,2)\delta _{\sigma
,\sigma ^{\prime }}\left\langle T_{\tau }c_{\sigma }^{\dagger }(1)c_{\sigma
}(1)\right\rangle }{\left\langle T_{\tau }c_{\sigma }^{\dagger }(1)c_{\sigma
}(1)\right\rangle \left\langle T_{\tau }c_{\sigma ^{\prime }}^{\dagger
}(2)c_{\sigma ^{\prime }}(2)\right\rangle }\equiv \frac{\left\langle
n_{\sigma }\left( 1\right) n_{\sigma ^{\prime }}\left( 2\right)
\right\rangle -\delta (1,2)\delta _{\sigma ,\sigma ^{\prime }}\left\langle
n_{\sigma }\left( 1\right) \right\rangle }{\left\langle n_{\sigma }\left(
1\right) \right\rangle \left\langle n_{\sigma ^{\prime }}\left( 2\right)
\right\rangle }
\end{equation}%
\end{widetext}%
where $\left\langle n_{\sigma }\left( 1\right) n_{\sigma ^{\prime }}\left( 2\right)
\right\rangle$ is the density-density correlation function. In this last formula it is
assumed that $\tau _{1}=\tau _{2}$. With this procedure, the four point
function $\left\langle T_{\tau }c_{\tilde{\sigma}}^{\dagger }(1)c_{\tilde{%
\sigma}}(1)c_{\sigma }(1)c_{\sigma }^{\dagger }(\bar{3})\right\rangle $
appearing in the definition of the self-energy Eq.(\ref{Self_exact}) is
factorized \`{a} la Hartree-Fock everywhere \cite{Allen:2003} except when
the point $\overline{3}$ is equal to $1^{+}$, in which case there is no
approximation involved. The Fock contribution from the $V$ term is discussed
in Appendix A. It gives a very small contribution in the regime studied in
the present paper. We caution the reader that the above-mentioned
factorization is not exactly the one which is used in the STLS
approximation. Further factorizations and additional details
will be discussed in the following sections.

We want to calculate the spin and charge response functions. These can be
obtained from the first functional derivative of the Green function with
respect to the external source field. Taking the functional derivative on
both sides of the identity $G_{\sigma }(1,\bar{3})G_{\sigma ^{\prime }}^{-1}(%
\bar{3},2)=\delta _{\sigma \sigma ^{\prime }}\delta (1,2)$ and using the
Dyson equation Eq.~(\ref{Dyson}), we obtain the exact result%
\begin{widetext}%
\begin{align}
\Pi _{\sigma \sigma ^{\prime }}(1,2;3,3)& \equiv -\frac{\delta G_{\sigma
}(1,2)}{\delta \phi _{\sigma ^{\prime }}(3,3)}=G_{\sigma }(1,\bar{4})\frac{%
\delta G_{\sigma }^{-1}(\bar{4},\bar{5})}{\delta \phi _{\sigma ^{\prime
}}(3,3)}G_{\sigma }(\bar{5},2)  \notag \\
& =-\delta _{\sigma \sigma ^{\prime }}G_{\sigma }(1,3)G_{\sigma
}(3,2)-\sum_{\sigma ^{\prime \prime }}G_{\sigma }(1,\bar{4})\frac{\delta
\Sigma _{\sigma }(\bar{4},\bar{5})}{\delta G_{\sigma ^{\prime \prime }}(\bar{%
6},\bar{7})}\frac{\delta G_{\sigma ^{\prime \prime }}(\bar{6},\bar{7})}{%
\delta \phi _{\sigma ^{\prime }}(3,3)}G_{\sigma }(\bar{5},2).  \label{Pi}
\end{align}%
In turn, the first functional derivative of the self-energy respect to the
Green function can be evaluated from our approximate expression for the
self-energy Eq.~(\ref{Self_approx})
\begin{align}
\frac{\delta \Sigma _{\sigma }(4,5)}{\delta G_{\sigma ^{\prime \prime }}(6,7)%
}& =U\delta _{\tilde{\sigma}\sigma ^{\prime \prime }}\delta (4,5)\delta
(4,6)\delta (5,7)g_{\sigma \tilde{\sigma}}(4,4)+V\sum_{a}\delta (4,5)\delta
(4+a,6)\delta (5+a,7)g_{\sigma \sigma ^{\prime \prime }}(4,4+a)  \notag
\label{dsdg} \\
& +U\delta (4,5)G_{\tilde{\sigma}}(4,4^{+})\frac{\delta g_{\sigma \tilde{%
\sigma}}(4,4)}{\delta G_{\sigma ^{\prime \prime }}(6,7)}+V\delta
(4,5)\sum_{\sigma ^{\prime \prime \prime },a}G_{\sigma ^{\prime \prime
\prime }}(4,4^{+})\frac{\delta g_{\sigma \sigma ^{\prime \prime \prime
}}(4,4+a)}{\delta G_{\sigma ^{\prime \prime }}(6,7)}.
\end{align}%
\end{widetext}%
The functional derivative of the pair correlation function with respect to
the Green function is a three-body (six-point) correlation function that is
not known. For the standard Hubbard model, it was shown that the unknown
functional derivative (third term in the above equation) appears only in the
charge response function. The authors in Ref.~(\onlinecite{Vilk1, Allen:2003})
approximated this functional by a constant whose value was obtained by
enforcing the Pauli principle expressed as a sum-rule on spin and charge
correlation functions. In our case, two other unknown functionals that come
from the last term in the above equation, appear in both the charge and spin
response functions. We assume, and confirm with the numerical results of the
following section, that these two unknown functionals do not give important
contributions as long as $4V$ is small compared to the bandwidth. Their contribution
becomes more significant as $V$ increases. By approximating the two unknown functions by
two different constants, it should be possible to obtain them by using two extra sum-rules,
such as the compressibility and spin susceptibility sum-rules. We leave this for future
work and, at this point, we simply drop the $\delta g_{\sigma \sigma
^{\prime \prime \prime }}(4,4+a)/\delta G_{\sigma ^{\prime \prime }}(6,7)$
term in the last line.

The spin and charge part of $\Pi $ can now be obtained by combining Eqs.~(%
\ref{Pi}) and (\ref{dsdg}) in the form%
\begin{widetext}%
\begin{align}
\Pi _{cc,ss}(1,2;3,3)\equiv&\sum_{\sigma\sigma^{\prime}}(\sigma\sigma^{\prime})
\Pi_{\sigma\sigma^{\prime}}(1,2;3,3)= 2\left[ \Pi _{\sigma \sigma }(1,2;3,3)\pm \Pi
_{\sigma \tilde{\sigma}}(1,2;3,3)\right]  \notag \\
=& -2G_{\sigma }(1,3)G_{\sigma }(3,2)
 -2UG_{\sigma }(1,\bar{4})G_{\sigma }(\bar{4},2)g_{\sigma \tilde{\sigma}}(%
\bar{4},\bar{4})\left[ \frac{\delta G_{\sigma }(\bar{4},\bar{4}^{+})}{\delta
\phi _{\tilde{\sigma}}(3,3)}\pm \frac{\delta G_{\sigma }(\bar{4},\bar{4}^{+})%
}{\delta \phi _{\sigma }(3,3)}\right]  \notag \\
& -2VG_{\sigma }(1,\bar{4})G_{\sigma }(\bar{4},2)\sum_{\sigma ^{\prime
},a}g_{\sigma \sigma ^{\prime }}(\bar{4},\bar{4}+a)\left[ \frac{\delta
G_{\sigma }(\bar{4}+a,\bar{4}+a^{+})}{\delta \phi _{\tilde{\sigma}}(3,3)}\pm
\frac{\delta G_{\sigma }(\bar{4}+a,\bar{4}+a^{+})}{\delta \phi _{\sigma
}(3,3)}\right]  \notag \\
& -2UG_{\sigma }(1,\bar{4})G_{\sigma }(\bar{4},2)G_{\tilde{\sigma}}(\bar{4},%
\bar{4})\left[ \frac{\delta g_{\sigma \tilde{\sigma}}(\bar{4},\bar{4})}{%
\delta \phi _{\sigma }(3,3)}\pm \frac{\delta g_{\sigma \tilde{\sigma}}(\bar{4%
},\bar{4})}{\delta \phi _{\tilde{\sigma}}(3,3)}\right].  \label{Piccss}
\end{align}%
\end{widetext}%
where $\sigma=\pm$ (the third identity in the above equation is valid for
a spin-unpolarized system).

To obtain the response functions of the system, we set the external
potential to zero. When the minus sign (corresponding to the spin response
function) is chosen in the last term, it drops out by rotational invariance
in zero source field \cite{Vilk2}. For the plus sign (corresponding to the
charge response function), we assume that the functional derivative of the
pair correlation respect to the density is a constant (after using an extra
chain rule in the above equation). The final form of the charge and spin
response functions, or equivalently susceptibilities, in Fourier space then
have the following forms
\begin{equation}
\chi _{ss}(\mathbf{q},\omega _{n})=\frac{\chi ^{0}(\mathbf{q},\omega _{n})}{%
1-\frac{\chi ^{0}(\mathbf{q},\omega _{n})}{2}U_{ss}(\mathbf{q})}
\label{chi_ss}
\end{equation}%
\begin{equation}
\chi _{cc}(\mathbf{q},\omega _{n})=\frac{\chi ^{0}(\mathbf{q},\omega _{n})}{%
1+\frac{\chi ^{0}(\mathbf{q},\omega _{n})}{2}U_{cc}(\mathbf{q})}
\label{chi_cc}
\end{equation}%
where
\begin{equation}
U_{ss}(\mathbf{q})= Ug_{\sigma \tilde{\sigma}}(0)-4Vg_{ss}(a)
\sum_{\alpha }\cos (q_{\alpha }a),
\end{equation}
\begin{equation}
U_{cc}(\mathbf{q})=U\left(
g_{\sigma \tilde{\sigma}}(0)+ng_{\sigma \tilde{\sigma}}^{\prime }(0)\right)
+4Vg_{cc}(a)\sum_{\alpha }\cos (q_{\alpha }a)
\end{equation}
and $\omega _{n}=2\pi nT$ is the Matsubara frequency. One can easily see
the similarity of the above equations with what we had in Refs. \cite{Vilk1,Vilk2}
and also find out how easily these equations can be extended to a system with longer
range of interaction and also to a system with many bands.
Finally  the index $\alpha $ takes the values $\alpha =1..D$, $D$ being the
dimension of the system, while the equal-time charge and spin pair correlation
functions are defined by  $g_{cc,ss}\equiv\sum_{\sigma\sigma^{\prime}}(\sigma\sigma^{\prime})
n_\sigma n_{\sigma'}g_{\sigma\sigma^{\prime}}/n^{2}$
(or simply $g_{cc,ss}(\mathbf{r})=(g_{\sigma \sigma }(\mathbf{r})\pm g_{\sigma
\tilde{\sigma}}(\mathbf{r}))/2$ for a spin-unpolarized system)
with $\left\vert \mathbf{r}\right\vert =0$
or $a$ and $\chi ^{0}(\mathbf{q},\omega _{n})$ is the response function for
non-interacting electrons given by
\begin{equation}
\chi ^{0}(\mathbf{q},\omega _{n})=\int_{BZ}\frac{d\mathbf{p}}{\nu }\frac{%
f^{0}(\mathbf{p}+\frac{\mathbf{q}}{2})-f^{0}(\mathbf{p}-\frac{\mathbf{q}}{2})%
}{i\omega _{n}-\epsilon _{\mathbf{p}+\mathbf{q}/2}+\epsilon _{\mathbf{p}-%
\mathbf{q}/2}}.
\end{equation}%
In the above formula $\nu $ is the volume of the Brillouin zone (BZ), $f^{0}(%
\mathbf{q})=1/[1+\exp ((\epsilon _{q}-\mu )/T)]$ is the Fermi-Dirac
distribution function, and $\epsilon _{q}=-2t\sum_{\alpha }\cos (q_{\alpha
}a)$ is the non-interacting particle dispersion relation. The pair
correlation functions are related to the static structure functions by
\begin{equation}
g_{cc}(\mathbf{r})=1+\frac{1}{n}\int_{BZ}\frac{d\mathbf{q}}{\nu }[S_{cc}(%
\mathbf{q})-1]\exp (i\mathbf{q}\cdot \mathbf{r})  \label{gcc}
\end{equation}%
\begin{equation}
g_{ss}(\mathbf{r})=\frac{1}{n}\int_{BZ}\frac{d\mathbf{q}}{\nu }[S_{ss}(%
\mathbf{q})-1]\exp (i\mathbf{q}\cdot \mathbf{r})  \label{gss}
\end{equation}%
where $S_{cc,ss}\equiv\sum_{\sigma\sigma^{\prime}}(\sigma\sigma^{\prime})
\sqrt{n_\sigma n_{\sigma'}}S_{\sigma\sigma^{\prime}}/n$
(or simply $S_{cc,ss}(\mathbf{r})\equiv S_{\sigma \sigma }(\mathbf{r})\pm S_{\sigma
\tilde{\sigma}}(\mathbf{r})$ for a spin-unpolarized system)
are respectively the spin and charge static 
structure factors, with the spin-resolved static structure
factor defined by $S_{\sigma \sigma ^{\prime }}(\mathbf{r})\equiv
\left[ \left\langle n_{\sigma }(\mathbf{0})n_{\sigma ^{\prime }}(\mathbf{r}%
)\right\rangle /\sqrt{n_\sigma n_{\sigma'}}\right]-\sqrt{n_\sigma n_{\sigma'}}$
and the Fourier transforms by,%
\begin{equation}
\sum_{\mathbf{l}}e^{i\mathbf{p}\cdot \mathbf{r}_{\mathbf{l}}}\equiv \nu
\delta (\mathbf{p}),\;\;\;\;\;\int_{BZ}\frac{d\mathbf{p}}{\nu }e^{i\mathbf{p}%
\cdot \mathbf{r}_{\mathbf{l}}}\equiv \delta _{\mathbf{r}_{\mathbf{l}},0}.
\label{FT}
\end{equation}%
Finally the static structure factors are connected to the response functions
by the fluctuation-dissipation theorem
\begin{equation}
S_{\sigma\sigma'}(\mathbf{q})=\frac{T}{\sqrt{n_\sigma n_{\sigma'}}}
\sum_{\omega _{n}}\chi _{\sigma\sigma'}(\mathbf{q},\omega _{n}).
\label{FDT}
\end{equation}

The above equations [Eqs. (\ref{chi_cc})-(\ref{FDT})] form eight relations
containing nine unknowns. The extra unknown can be fixed using the Pauli
principle, namely $g_{\sigma \sigma }(0)=0$ or $g_{cc}(0)=-g_{ss}(0)$.

To conclude, note that the RPA approximation on the nearest-neighbor
interaction $V$ can be simply recovered by setting $g_{cc}(1)=1$ and $%
g_{ss}(1)=0$ which means that the RPA does not give any extra correction to
the spin response function of the extended Hubbard model. This is a
consequence of the fact that the different spin components (spin parallel
and anti-parallel) of the off-site interaction are identical in the original
Hamiltonian. We will see that this is in contradiction with QMC
calculations. Thus, our generalization of TPSC does take into account
important non-perturbative corrections. Finally, at this level of approximation
the self-energy Eq.~(\ref{Self_approx}) is a constant. As in the original TPSC
approach \cite{Vilk3, Moukouri:2000, Allen:2003}, we can perform a second
step to improve our approximation for the self-energy. This is discussed in
Appendix \ref{App1}.

\subsection{Wigner distribution function approach}

In this section, we present the approach for obtaining the structure factors
that parallels that of STLS \cite{Singwi}. It is based on the equation of
motion for the Wigner distribution function. We just show the calculation
for the ordinary Hubbard model $\left( V=0\right) $ in order to shorten the
length of the equations. This will suffice to demonstrate the difference
between TPSC and STLS.

The one- and two-body Wigner distribution function (OBWDF and TBWDF) are
defined by,
\begin{equation}
f_{\mathbf{i}\sigma }(\mathbf{p},\tau )=\sum_{\mathbf{l}}e^{i\mathbf{p}\cdot %
\mathbf{r}_{\mathbf{l}}}\left\langle c_{\mathbf{i}+\mathbf{l}/2,\sigma%
}^{\dagger }c_{\mathbf{i}-\mathbf{l}/2,\sigma }\right\rangle%
\end{equation}
\begin{widetext}%
\begin{equation}
f_{\mathbf{i}\mathbf{i}^{\prime }\sigma \sigma ^{\prime }}(\mathbf{p},%
\mathbf{p}^{\prime },\tau )=\sum_{\mathbf{l},\mathbf{l}^{\prime }}e^{i%
\mathbf{p}\cdot \mathbf{r}_{\mathbf{l}}}e^{i\mathbf{p}^{\prime }\cdot
\mathbf{r}_{\mathbf{l}^{\prime }}}\left\langle c_{\mathbf{i}+\mathbf{l}%
/2,\sigma }^{\dagger }c_{\mathbf{i}-\mathbf{l}/2,\sigma }c_{\mathbf{i}%
^{\prime }+\mathbf{l}^{\prime }/2,\sigma ^{\prime }}^{\dagger }c_{\mathbf{i}%
^{\prime }-\mathbf{l}^{\prime }/2,\sigma ^{\prime }}\right\rangle .
\end{equation}%
\end{widetext}%
One should notice that the operators in the above equations act on some
lattice sites which do not exist in the real system. Knowing that the Winger
distribution functions (WDFs) are not a real physical functions we define them
in this manner for the sake of simplicity in the notation. 
One can also define the WDFs in term of the operators which
just act on the real lattice sites, but that makes the formalism a bit more tedious.

The density of particles of spin $\sigma $ at position $\mathbf{i}$ is
related to the OBWDF by
\begin{equation}
n_{\mathbf{i}\sigma }(\tau )=\int_{BZ}\frac{d\mathbf{p}}{\nu }f_{\mathbf{i}%
\sigma }(\mathbf{p},\tau )  \label{density}
\end{equation}%
with the same definitions of Fourier transforms as above, Eq.~(\ref{FT}).

We first need to write the equation of motion for the operator $c_{\mathbf{i}%
,\sigma }$ to obtain the equation of motion for the OBWDF later on. This
equation, after a bit of algebra, can be written as,
\begin{equation}
-\frac{\partial c_{\mathbf{i},\sigma }}{\partial \tau }=\left[ H,c_{\mathbf{i%
},\sigma }\right] =-t\delta _{\mathbf{i}}^{2}(c_{\mathbf{i},\sigma })+Uc_{%
\mathbf{i},\sigma }c_{\mathbf{i},\tilde{\sigma}}^{\dagger }c_{\mathbf{i},%
\tilde{\sigma}}+V_{\mathbf{i},\sigma }^{ext}c_{\mathbf{i},\sigma }
\label{EOM_c}
\end{equation}%
where $\delta _{\mathbf{i}}^{2}c_{\mathbf{i},\sigma }=\sum_{\left\langle
\mathbf{j}\right\rangle _{\mathbf{i}}}c_{\mathbf{j},\sigma }$ and ${%
\left\langle \mathbf{j}\right\rangle _{\mathbf{i}}}$ means that the sum runs
over all nearest-neighbors of site $\mathbf{i}$. Using the above equation,
one can write the equation of motion for the OBWDF,%
\begin{widetext}
\begin{align}
-\frac{\partial f_{\mathbf{i},\sigma }(\mathbf{p},\tau )}{\partial \tau }&
=-2it\sum_{\alpha }\sin (q_{\alpha })(\delta _{\mathbf{i}})_{\alpha }f_{%
\mathbf{i},\sigma }(\mathbf{p},\tau )  \notag  \label{emf} \\
& -\sum_{\mathbf{l}}\int_{BZ}\frac{d\mathbf{p}_{1}}{\nu }e^{i(\mathbf{p}-%
\mathbf{p}_{1})\cdot \mathbf{r}_{\mathbf{l}}}(V_{\mathbf{i}+\mathbf{l}%
/2,\sigma }^{ext}-V_{\mathbf{i}-\mathbf{l}/2,\sigma }^{ext})f_{\mathbf{i}%
,\sigma }(\mathbf{p}_{1},\tau )  \notag \\
& -U\sum_{\mathbf{l},\mathbf{i}^{\prime }}\int_{BZ}\frac{d\mathbf{p}_{1}}{%
\nu }\int_{BZ}\frac{d\mathbf{p}_{2}}{\nu }e^{i(\mathbf{p}-\mathbf{p}%
_{1})\cdot \mathbf{r}_{\mathbf{l}}}(\delta _{\mathbf{i}-\mathbf{i}^{\prime },%
\mathbf{l}/2}-\delta _{\mathbf{i}-\mathbf{i}^{\prime },-\mathbf{l}/2})f_{%
\mathbf{i}\mathbf{i}^{\prime }\sigma \tilde{\sigma}}(\mathbf{p}_{1},\mathbf{p%
}_{2},\tau ).
\end{align}%
To derive the first term on the right-hand side of the above equation, we
used the following identity, that we prove in one-dimension only (for the
sake of simplicity of notation):

\begin{align}
& \delta _{\mathbf{i}}^{2}(c_{i+l/2,\sigma }^{\dagger })c_{i-l/2,\sigma
}-c_{i+l/2,\sigma }^{\dagger }\delta _{\mathbf{i}}^{2}(c_{i-l/2,\sigma })
\notag \\
& =c_{i+l/2+1,\sigma }^{\dagger }c_{i-l/2,\sigma }-c_{i+l/2,\sigma
}^{\dagger }c_{i-l/2-1,\sigma }+c_{i+l/2-1,\sigma }^{\dagger
}c_{i-l/2,\sigma }-c_{i+l/2,\sigma }^{\dagger }c_{i-l/2+1,\sigma }  \notag \\
& =\delta _{i}(c_{i+l/2+1/2,\sigma }^{\dagger }c_{i-l/2-1/2,\sigma })-\delta
_{i}(c_{i+l/2-1/2,\sigma }^{\dagger }c_{i-l/2+1/2,\sigma })  \notag \\
& =\delta _{i}\delta _{2l}(c_{i+l/2,\sigma }^{\dagger }c_{i-l/2,\sigma }),
\end{align}%
where $\delta _{i}h_{i,\sigma }=h_{i+1/2,\sigma }-h_{i-1/2,\sigma }$ and $%
\delta _{2i}h_{i,\sigma }=h_{i+1,\sigma }-h_{i-1,\sigma }$ where $h_i$ is a
general function of the operators $c_i$ and $c_i^\dagger$, such as
$h_i=c_{i+j}^\dagger c_{i+k}$ in which $j$ and $k$ are arbitrary numbers.

The TBWDF appears in the equation of motion for the OBWDF Eq.~(\ref{emf}),
which means that we have to make an approximation in order to obtain a
closed set of equations. Proceeding by analogy with the previous section, we
factor the TBWDF as follows,%
\begin{align}
f_{\mathbf{i}\mathbf{i}^{\prime }\sigma \tilde{\sigma}}(\mathbf{p},\mathbf{p}%
^{\prime },\tau )& \approx f_{\mathbf{i}\sigma }(\mathbf{p},\tau )f_{\mathbf{%
i}^{\prime }\tilde{\sigma}}(\mathbf{p}^{\prime },\tau )g_{\sigma \tilde{%
\sigma}}(\mathbf{i}^{\prime },\mathbf{i}^{\prime },\tau )
\label{Factorization_TPSC} \\
& \approx f_{\sigma }(\mathbf{p})f_{\tilde{\sigma}}(\mathbf{p}^{\prime
})g_{\sigma \tilde{\sigma}}(0)+f_{\sigma }(\mathbf{p})\bar{f}_{\mathbf{i}%
^{\prime }\tilde{\sigma}}(\mathbf{p}^{\prime },\tau )g_{\sigma \tilde{\sigma}%
}(0)+\bar{f}_{\mathbf{i}\sigma }(\mathbf{p},\tau )f_{\tilde{\sigma}}(\mathbf{%
p}^{\prime })g_{\sigma \tilde{\sigma}}(0)  \notag \\
& +f_{\sigma }(\mathbf{p})f_{\tilde{\sigma}}(\mathbf{p}^{\prime })\sum_{%
\mathbf{j},\sigma ^{\prime }}\frac{\partial g_{\sigma \tilde{\sigma}}(%
\mathbf{i}^{\prime },\mathbf{i}^{\prime },\tau )}{\partial n_{\mathbf{j}%
,\sigma ^{\prime }}(\tau )}\bar{n}_{\mathbf{j},\sigma ^{\prime }}(\tau )
\end{align}%
\end{widetext}%
where we define $f_{\mathbf{i}\sigma }(\mathbf{p},\tau )\cong f_{\sigma }(%
\mathbf{p})+\bar{f}_{\mathbf{i}\sigma }(\mathbf{p},\tau )$ with $\bar{f}_{%
\mathbf{i}\sigma }(\mathbf{p},\tau )$ the deviation of the OBWDF from its
average value due to presence of the external potential. In addition we
assume that the external potential is weak enough that we can keep only the
first term in the functional Taylor expansion of $g_{\sigma \tilde{\sigma}}(%
\mathbf{i}^{\prime },\mathbf{i}^{\prime },\tau ),$ which means that $\bar{f}%
_{\mathbf{i}\sigma }(\mathbf{p},\tau )$ and $\bar{n}_{\mathbf{j},\sigma
^{\prime }}(\tau )$ are small. The first and third terms of the above
equation do not contribute to the final form of the equation of motion for
the OBWDF, Eq.~(\ref{emf}). The functional dependence of $g_{\sigma \tilde{%
\sigma}}(\mathbf{i},\mathbf{i}^{\prime },\tau )$ on $n_{\mathbf{j},\sigma
^{\prime }}(\tau )$ again appears in the above equation. We use the
well-known local approximation $\partial g_{\sigma \tilde{\sigma}}(\mathbf{i}%
,\mathbf{i},\tau )/\partial n_{\mathbf{j},\sigma ^{\prime }}(\tau )=\delta
_{i,j}\partial g_{\sigma \tilde{\sigma}}(0)/\partial n_{\sigma ^{\prime }}$
where $n_{\sigma }$ is the average number of particles per site with spin $%
\sigma $. The final non-zero contribution from the above approximation for
the TBWDF to the equation of the motion finally takes the following form,%
\begin{widetext}
\begin{equation}
K_{\mathbf{i}\mathbf{i}^{\prime }\sigma \sigma ^{\prime }}(\mathbf{p},%
\mathbf{p}^{\prime },\tau )=f_{\sigma }(\mathbf{p})\bar{f}_{\mathbf{i}%
^{\prime }\tilde{\sigma}}(\mathbf{p}^{\prime },\omega )g_{\sigma \tilde{%
\sigma}}(0)+f_{\sigma }(\mathbf{p})f_{\tilde{\sigma}}(\mathbf{p}^{\prime
})\sum_{\sigma ^{\prime }}\frac{\partial g_{\sigma \tilde{\sigma}}(0)}{%
\partial n_{\sigma ^{\prime }}}\bar{n}_{\mathbf{i}^{\prime },\sigma ^{\prime
}}(\tau ).
\end{equation}%
The exact form of $f_{\sigma }(\mathbf{p})$ is not known but in first approximation
it is reasonable to replace it by the Fermi-Dirac function $%
f_{\sigma }^{0}(\mathbf{p})$. The final equation for $\bar{f}_{\mathbf{i}%
\sigma }(\mathbf{p},\tau )$ in Fourier space can finally be written as
\begin{align}
\lbrack \omega _{n}& -4t\sum_{\alpha }\sin (\frac{q_{\alpha }}{2})\sin
(p_{\alpha })]\bar{f}_{\sigma }(\mathbf{q},\mathbf{p},\omega
_{n})=[f_{\sigma }^{0}(\mathbf{p}+\frac{\mathbf{q}}{2})-f_{\sigma }^{0}(%
\mathbf{p}-\frac{\mathbf{q}}{2})]V_{\sigma }^{ext}(\mathbf{q},\omega _{n})
\notag \\
& +U\sum_{\mathbf{l}}\int_{BZ}\frac{d\mathbf{p}_{1}}{\nu }\int_{BZ}\frac{d%
\mathbf{p}_{2}}{\nu }e^{i(\mathbf{p}-\mathbf{p}_{1})\cdot \mathbf{r}_{%
\mathbf{l}}}[e^{i\frac{\mathbf{q}}{2}\cdot \mathbf{r}_{\mathbf{l}}}-e^{-i%
\frac{\mathbf{q}}{2}\cdot \mathbf{r}_{\mathbf{l}}}]K_{\sigma \sigma ^{\prime
}}(\mathbf{q},\mathbf{p},\mathbf{p}^{\prime },\omega _{n}).
\end{align}%
Now we can perform an integral over $\mathbf{p}$ to obtain an equation for
the density $\bar{n}_{\sigma }(\mathbf{q},\omega _{n})$. The final form of
this equation is
\begin{equation}
\bar{n}_{\sigma }(\mathbf{q},\omega _{n})=\chi _{\sigma }^{0}(\mathbf{q}%
,\omega _{n})V_{\sigma }^{ext}(\mathbf{q},\omega _{n})+U\chi _{\sigma }^{0}(%
\mathbf{q},\omega _{n})[\bar{n}_{\tilde{\sigma}}(\mathbf{q},\omega
_{n})g_{\sigma \tilde{\sigma}}(0)+n_{\sigma }\sum_{\sigma ^{\prime }}\frac{%
\partial g_{\sigma \tilde{\sigma}}(0)}{\partial n_{\sigma ^{\prime }}}\bar{n}%
_{\sigma ^{\prime }}(\mathbf{q},\omega _{n})]  \label{rho}
\end{equation}%
\end{widetext}%
where $\chi _{\sigma }^{0}(\mathbf{q},\omega )$ is given by,
\begin{equation}
\chi _{\sigma }^{0}(\mathbf{q},\omega _{n})=\int_{BZ}\frac{d\mathbf{p}}{\nu }%
\frac{f_{\sigma }^{0}(\mathbf{p}+\frac{\mathbf{q}}{2})-f_{\sigma }^{0}(%
\mathbf{p}-\frac{\mathbf{q}}{2})}{i\omega _{n}+\epsilon _{\mathbf{p}+\mathbf{%
q}/2}-\epsilon _{\mathbf{p}-\mathbf{q}/2}}.
\end{equation}%
One can invert the equation for the change in density Eq.~(\ref{rho}) in
order to obtain the density in terms of the external potential to extract
the susceptibility,%
\begin{equation*}
\bar{n}_{\sigma }(\mathbf{q},\omega _{n})=\chi _{\mathrm{cc}}(\mathbf{q}%
,\omega _{n})V_{\mathrm{cc}}^{ext}(\mathbf{q},\omega _{n})+\chi _{\mathrm{ss}%
}(\mathbf{q},\omega _{n})V_{\mathrm{ss}}^{ext}(\mathbf{q},\omega _{n})
\end{equation*}%
where $V_{\mathrm{cc},\mathrm{ss}}^{ext}(\mathbf{q},\omega _{n})=[V_{\sigma
}^{ext}(\mathbf{q},\omega _{n})\pm V_{\tilde{\sigma}}^{ext}(\mathbf{q}%
,\omega _{n})]/2$, and the coefficients of the external potential are the response
functions, which are given by following formulae,
\begin{equation}
\chi _{\mathrm{cc},\mathrm{ss}}(\mathbf{q},\omega _{n})=\frac{\chi ^{0}(%
\mathbf{q},\omega _{n})}{1\mp \frac{U_{\mathrm{cc},\mathrm{ss}}}{2}\chi ^{0}(%
\mathbf{q},\omega _{n})}  \label{chi_ccss}
\end{equation}%
where $\chi ^{0}(\mathbf{q},\omega _{n})=2\left( \chi _{\sigma }^{0}(\mathbf{%
q},\omega _{n})+\chi _{\tilde{\sigma}}^{0}(\mathbf{q},\omega _{n})\right) $,
$U_{\mathrm{ss}}=Ug_{\sigma \tilde{\sigma}}(0)$ and $U_{\mathrm{cc}%
}=U[g_{\sigma \tilde{\sigma}}(0)+n\partial g_{\sigma \tilde{\sigma}%
}(0)/\partial n]$.

Eqs.~(\ref{chi_ccss}) are the same as Eqs.~(\ref{chi_cc}) and Eq.~(\ref%
{chi_ss}) when we set $V=0$. The extension of the above equations to the
case $V\neq 0$ is straightforward and leads to exactly the same result as in
the previous section. In the present case, the derivative of the pair
correlation function with respect to the density can be evaluated if one
wishes, but its contribution is not big enough to reproduce the QMC results
as we will show in the next section. This problem is known in the context of
the electron liquid \cite{Vashishta}. The authors add a extra unknown
multiplier constant and fix it by the compressibility sum-rule. If they had
used instead Pauli principle they would have recovered the TPSC equations.

\subsection{Comments on the STLS approximation}

\label{STLS}

We are now in a position to contrast the results of the above section with
the STLS approximation. For the sake of simplicity, it is preferable to
limit ourselves to the case $V=0$. The factorization of the TBWDF that leads
to the STLS approximation is given by 
\begin{equation}
f_{\mathbf{i}\mathbf{i}^{\prime }\sigma \tilde{\sigma}}(\mathbf{p},\mathbf{p}%
^{\prime },\tau )\approx f_{\mathbf{i}\sigma }(\mathbf{p},\tau )f_{\mathbf{i}%
^{\prime }\tilde{\sigma}}(\mathbf{p}^{\prime },\tau )g_{\sigma \tilde{\sigma}%
}(\mathbf{i},\mathbf{i}^{\prime },\tau ).  \label{Factorization_STLS}
\end{equation}%
This should be contrasted with the TPSC factorization appearing in Eq.~(\ref%
{Factorization_TPSC}) where the pair correlation function is taken
\textquotedblleft on-site\textquotedblright . At first glance the STLS
factorization looks more reasonable because, as far as the TBWDF is
concerned, the integral of the last formula with respect to $\mathbf{p}$ and 
$\mathbf{p}^{\prime }$ leads to the exact result 
\begin{equation}
\left\langle n_{\mathbf{i}\sigma }(\tau )n_{\mathbf{i}^{\prime }\sigma
^{\prime }}(\tau )\right\rangle =\left\langle n_{\mathbf{i}\sigma }(\tau
)\right\rangle \left\langle n_{\mathbf{i}^{\prime }\sigma ^{\prime }}(\tau
)\right\rangle g_{\sigma \tilde{\sigma}}(\mathbf{i},\mathbf{i}^{\prime
},\tau ).
\end{equation}%
However, one must recall that in the equations of motion, the TBWDF appears
weighted by the range-dependent potential appearing in the Hamiltonian. In
particular, the form $f_{\mathbf{i}\mathbf{i}^{\prime }\sigma \tilde{\sigma}%
}(\mathbf{p},\mathbf{p}^{\prime },\tau )$ is valid only for interactions
with a finite range. With a local interaction, three of the
creation-annihilation operators are at the same point, as can be seen in Eq.~%
(\ref{EOM_c}). The factorization that appears correct, as judged by
comparisons with QMC, is the one that takes the role of the potential into
account. In the case of the simple Hubbard model, the potential is local in
time and space so one needs a local factorization to model the interaction
terms a best as possible.

The formal STLS approximation can be obtained by replacing the above STLS
factorization Eq.~(\ref{Factorization_STLS}) in the equation of motion Eq.~%
(\ref{emf}) and then by repeating the same steps as above. We must also
ignore the functional derivative of the pair correlation function with
respect to the density to recover the simplest result. The final forms of the
response functions are given by
\begin{equation}
\chi _{\mathrm{cc,ss}}(\mathbf{q},\omega _{n})=\frac{\chi ^{0}(\mathbf{q}%
,\omega _{n})}{1\mp \frac{U}{2}[1-G_{\sigma \tilde{\sigma}}(\mathbf{q}%
,\omega _{n})]\chi ^{0}(\mathbf{q},\omega _{n})}
\end{equation}%
where $G_{\sigma \tilde{\sigma}}(\mathbf{q},\omega _{n})$ is the local field
factor for the qSTLS approximation. It can be written as 
\begin{equation}
G_{\sigma \tilde{\sigma}}(\mathbf{q},\omega _{n})=-\frac{2}{n}\int_{BZ}\frac{d%
\mathbf{k}}{\nu }S_{\sigma \tilde{\sigma}}(\mathbf{k}-\mathbf{q})\frac{\chi
_{\sigma }^{0}(\mathbf{q},\mathbf{k},\omega _{n})}{\chi _{\sigma }^{0}(%
\mathbf{q},\omega _{n})}
\end{equation}%
where $\chi _{\sigma }^{0}(\mathbf{q},\mathbf{k},\omega _{n})$ is the
inhomogeneous free response function 
\begin{equation}
\chi _{\sigma }^{0}(\mathbf{q},\mathbf{k},\omega _{n})=\int_{BZ}\frac{d%
\mathbf{p}}{\nu }\frac{f_{\sigma }^{0}(\mathbf{p}+\frac{\mathbf{k}}{2}%
)-f_{\sigma }^{0}(\mathbf{p}-\frac{\mathbf{k}}{2})}{\omega _{n}+\epsilon _{%
\mathbf{p}+\mathbf{q}/2}-\epsilon _{\mathbf{p}-\mathbf{q}/2}}.
\end{equation}%
The local field factor in the STLS approximation can be obtain by taking the
following limit \cite{Moudgil},%
\begin{widetext}
\begin{equation}
G_{\sigma \tilde{\sigma}}^{STLS}(\mathbf{q})=\lim_{\omega _{n}\rightarrow
\infty }G_{\sigma \tilde{\sigma}}(\mathbf{q},\omega _{n})=-\frac{2}{n}\int_{BZ}%
\frac{d\mathbf{k}}{\nu }S_{\sigma \tilde{\sigma}}(\mathbf{k}-\mathbf{q})\frac{%
\sum_{\alpha }\sin (\frac{\mathbf{q}_{\alpha }}{2})\sin (\frac{\mathbf{k}%
_{\alpha }}{2})}{\sum_{\alpha }\sin ^{2}(\frac{\mathbf{q}_{\alpha }}{2})}.
\end{equation}%
This integral can be simplified using $\mathbf{k}\rightarrow \mathbf{k}+%
\mathbf{q}$ so that the final result appears as $\mathbf{q}$ independent, 
\begin{equation}
G_{\sigma \tilde{\sigma}}^{STLS}=-\frac{2}{nD}\int_{BZ}\frac{d\mathbf{k}}{\nu }%
S_{\sigma \tilde{\sigma}}(\mathbf{k})\sum_{\alpha }\cos (\frac{\mathbf{k}%
_{\alpha }}{2}).
\end{equation}

\end{widetext}%

\section{Numerical results}

\label{sec4}

We now present our numerical results obtained from Eqs. (\ref{chi_cc})-(\ref%
{FDT}). Unless mentioned otherwise, we use $U=4$ and $n=1$ in all the
figures, in units where $\hbar =k_{B}=t=1$. We first present $V=0$ results
to contrast TPSC with other approaches and understand the source of the
differences, then we move to the more general case.

\subsection{TPSC, STLS and other approaches for $V=0.$}

\begin{figure}[tbp]
\begin{center}
\includegraphics[scale=0.5]{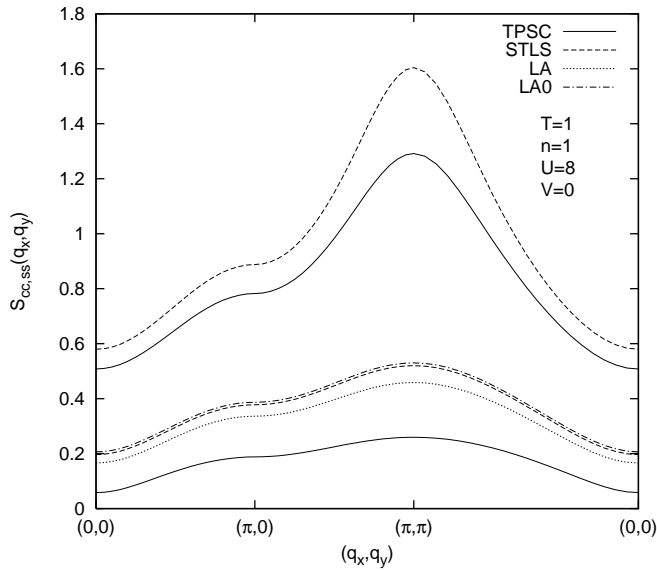}
\end{center}
\caption{The static structure factors for different methods at $U=8$, $n=1$
and $T=1$ as a function of momentum. The upper and lower curves are related
to spin and charge components respectively.}
\end{figure}

\begin{figure}[tbp]
\begin{center}
\includegraphics[scale=0.5]{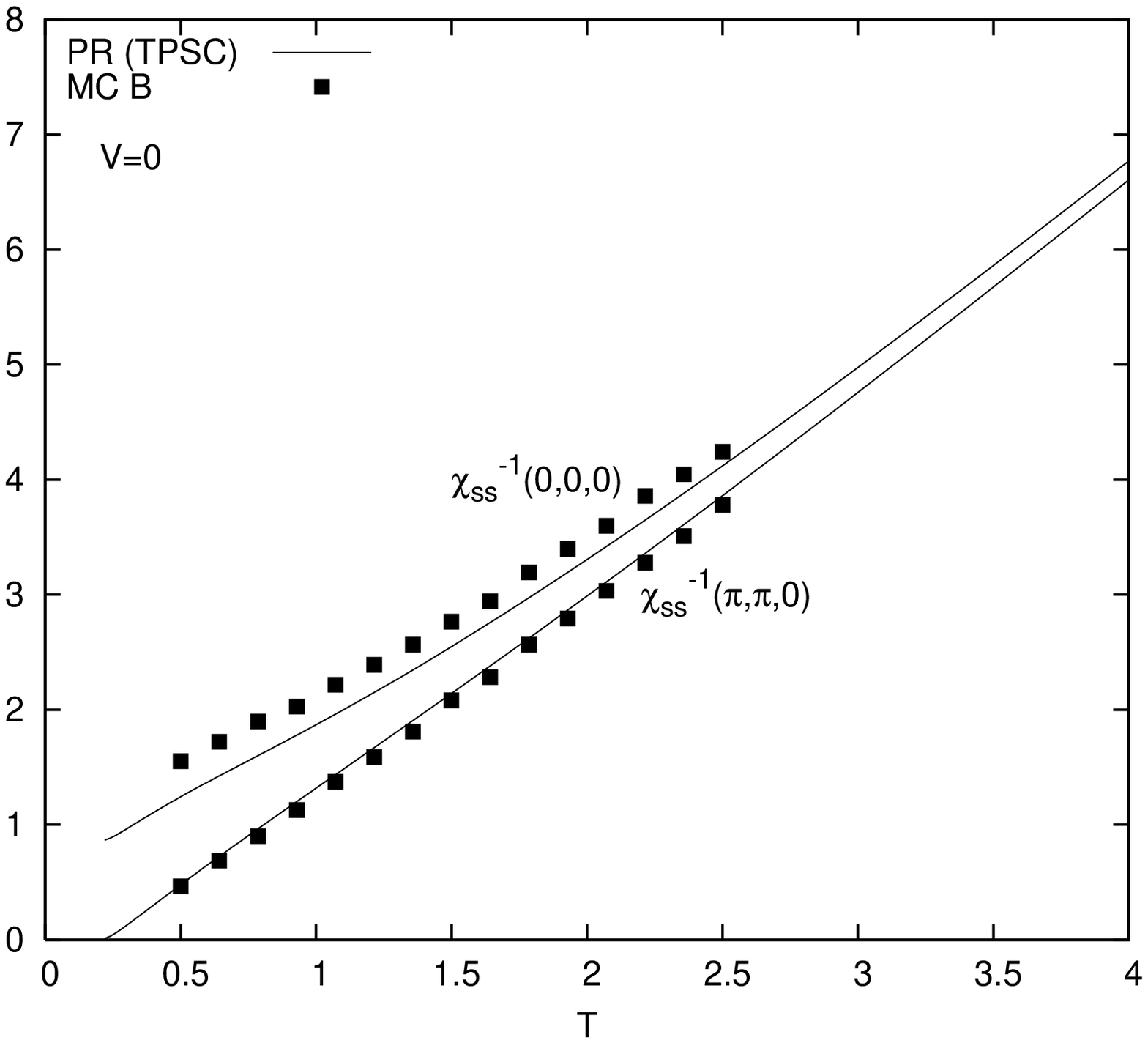}
\end{center}
\caption{The inverse of the spin part of the static response function at
specific momenta and $V=0,$ $U=4,$ $n=1$ as a function of temperature.
Symbols are QMC results extracted from \onlinecite{Zhang}.}
\end{figure}

In Fig.~1 we compare the static structure factors for different methods at $%
U=8$, $n=1$ and $T=1$. We take the TPSC results represented by the solid
lines in Fig.~1 (spin on top, charge on the bottom) as our reference.
Indeed, it was shown in great detail before \cite{Vilk1, Vilk2}, that the
TPSC values for the spin and charge structure factors agree very closely
with QMC calculations that are essentially exact within small statistical
uncertainties. However, TPSC is a weak to intermediate coupling method, so
it eventually fails for $U>8t$. Nevertheless, if one is not too close to
phase transitions, TPSC results for the spin structure factor are still
excellent at $U=8t$ while the results for the charge structure factor begin
to deviate from QMC because of the approximation involved in the evaluation
of the functional derivative. The STLS results, represented by the
long-dashed line in Fig.~1, deviate substantially from TPSC. The inaccuracy
of the STLS method for the Hubbard model comes from the fact that the
potential is local, so one should use the local pair correlation function to
correct the factorization in the equation of motion instead of the non-local
factoring used in the STLS approach, as discussed in Sec.~\ref{STLS}. In
addition, since both charge and spin structure factors for STLS are larger
than for TPSC, it is clear that STLS does not satisfy the Pauli principle $%
\left\langle n_{\sigma }^{2}\right\rangle =\left\langle n_{\sigma
}\right\rangle $ (or $g_{\sigma \sigma }(0)=0$), a key requirement for
electrons on a lattice at large density, where the probability of having two
electrons on the same site is large. For the charge structure factor, one
can compare two more methods with STLS and TPSC. The local approximation
(LA), represented by dots, consists in using for the effective interaction
in the charge channel Eq.~(\ref{chi_ccss}), $U_{\mathrm{cc}}=U[g_{\sigma 
\tilde{\sigma}}(0)+n\partial g_{\sigma \tilde{\sigma}}(0)/\partial n]$. We
call LA0, represented by dot-dashed line in Fig.~1, the approximation that
neglects completely $n\partial g_{\sigma \tilde{\sigma}}(0)/\partial n$. The
difference between STLS and LA0 is very small. However, the difference
between TPSC, LA, STLS (or LA0) is relatively large, demonstrating the
importance of the functional derivative in this range of physical
parameters. In the language of authors involved with the local
approximation, since LA does not provide a satisfactory result, the
multiplier factor in front of the derivative of the pair correlation
function respect to the density is necessary. When this unknown multiplier
is obtained from the Pauli sum-rule $g_{\sigma \sigma }(0)=0$ instead of
from the compressibility sum-rule, one recovers TPSC.

In TPSC, it suffices to know $g_{\sigma \tilde{\sigma}}(0)$ to obtain the
static spin structure factor. Could this quantity be determined from the
compressibility sum-rule instead of from the sum-rule relating $g_{\sigma 
\tilde{\sigma}}(0)$ to the integral of the structure factor? To answer this
question, we show in Fig.~2 the TPSC results for the inverse of the static $%
\left( \omega =0\right) $ spin response function (susceptibility) as a
function of temperature, again for the $V=0$ case, compared with QMC results
of Ref.~(\onlinecite{Zhang}). Both short and long wave-length limits show a
linear behavior in the intermediate and high temperature regimes, exhibiting
a Curie law. The deviations from the Curie law appear at low temperature in
both QMC and in TPSC. The agreement between TPSC and QMC is much better near
wave vector $\left( \pi ,\pi \right) $, even though the deviations are not
large even around $\left( 0,0\right) .$ Hence, the spin susceptibility
sum-rule is also very nearly satisfied with this method, meaning that the
momentum independent correction factor, which is given by $g_{\sigma \tilde{%
\sigma}}(0)$ in the effective interaction, corrects the result properly over
the entire Brillouin zone \cite{Nelisse:1999}. However, to use the spin
susceptibility sum-rule to fix the constant correction factor, one needs an
independent way to find the spin-susceptibility. Normally, the long
wave-length behavior of the spin (charge) response function is related to
the second derivative of the free energy with respect to magnetization $%
m=n_{\uparrow }-n_{\downarrow }$ (density $n$) which can then be computed
from the free-energy. In TPSC however, the free-energy requires further
study \cite{Roy:2005}. In addition, given the less accurate results
exhibited in Fig.~2 near $\left( 0,0\right) $, we consider it far more
preferable to use the original TPSC method where $g_{\sigma \tilde{\sigma}%
}(0)$ is determined by an integral over all wave vectors (Eqs. (\ref{gss})
and (\ref{FDT})) so as to satisfy the Pauli principle $g_{\sigma \sigma
}(0)=0$, which also involves a sum over all wave vectors.

\subsection{QMC vs generalization of TPSC for $V\neq 0$}

\begin{figure}[tbp]
\begin{center}
\includegraphics[scale=0.5]{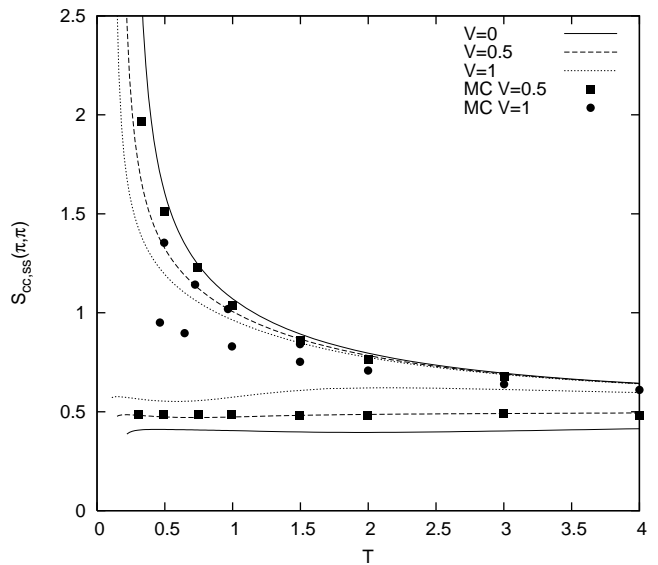}
\end{center}
\caption{The staggered static structure factors as a function of temperature
for $U=4,$ $n=1,$ $V=0,\;0.5\;\mathrm{and}\;1$. The dots are the QMC results
of Ref.~(\protect\cite{Zhang}) while lines are our results. The upper and lower
curves (dots) are related to the spin and charge components respectively.}
\end{figure}

\begin{figure}[tbp]
\begin{center}
\includegraphics[scale=0.5]{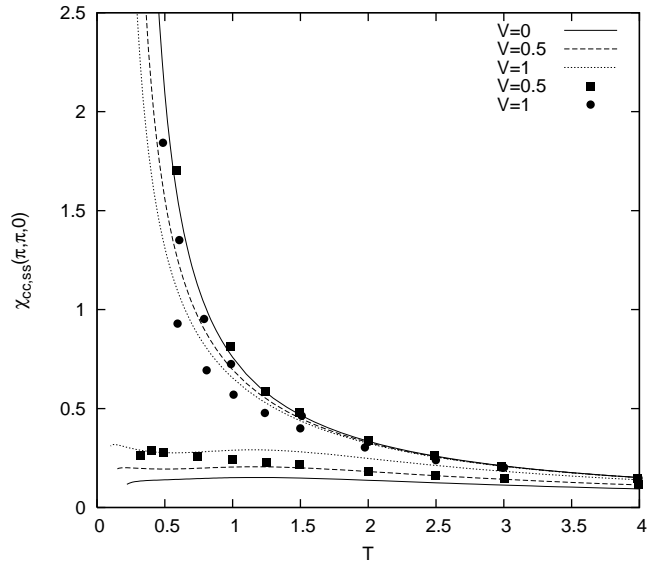}
\end{center}
\caption{The staggered static charge and spin susceptibilities as a function
of temperature using the same parameters and symbols as in Fig.~3.}
\end{figure}
To judge the accuracy of TPSC, we plot in Fig.~3 the staggered static
structure factors as a function of temperature, using QMC results of Ref.~(\onlinecite{Zhang}) 
as reference. The results for the charge structure factor
are all smaller than the corresponding results for the spin structure
factor. Our generalization for TPSC is plotted for $V=0,\;0.5\;$and$\;1$
while the QMC results, represented by symbols, are for $V=0.5\;$and$\;1$.
The figure clearly shows good agreement between our results and QMC in the
high temperature region or when $V=0.5$. The deviation becomes significant
when $V=1$ and the temperature is low. This suggests that the effect of the
functional derivative that we ignored in writing down Eqs.(\ref{chi_cc}) and
(\ref{chi_ss}) becomes important when when $4V$ is of the order of the bandwidth
(which is $8t$ here). Indeed the factor of four is necessary to account for the 
number of neighbors. The QMC data shows (not on the
figure) that the tendency to staggered spin order disappears around $%
V\approx 1.25$ while in our case it persists to a higher value $V\approx 2.$
This means that even though $g_{ss}(a)$ decreases with increasing $V,$ the
combination $Vg_{ss}(a)$ does not increase fast enough. The functional
derivative must become important to take this effect into account. Note that
in RPA, the spin structure factor is independent of $V$.

Finally, Fig.~4 shows the staggered static charge and spin susceptibilities
as a function of the temperature for the same parameters as the previous
figure. All features are similar to what was mentioned in Fig.~3, except
this fact that the correction factors do not correct the structure functions
and response functions in the same manner. This is due to this assumption that 
these functions are local in time and space which certainly fails in certain 
region of the parameters even in the case $V=0$.

\subsection{The effect of $V$}

\begin{figure}[tbp]
\begin{center}
\includegraphics[scale=0.5]{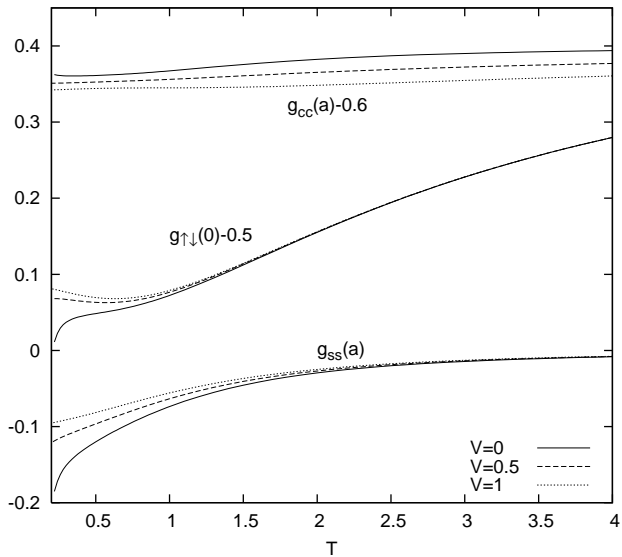}
\end{center}
\caption{The variation of $g_{\protect\sigma \tilde{\protect\sigma}}(0)$ and 
$g_{cc,ss}(a)$ as a function of temperature for $U=4,\;n=1$ and $%
V=0,\;0.5,\;1$.}
\end{figure}

\begin{figure}[tbp]
\begin{center}
\includegraphics[scale=0.5]{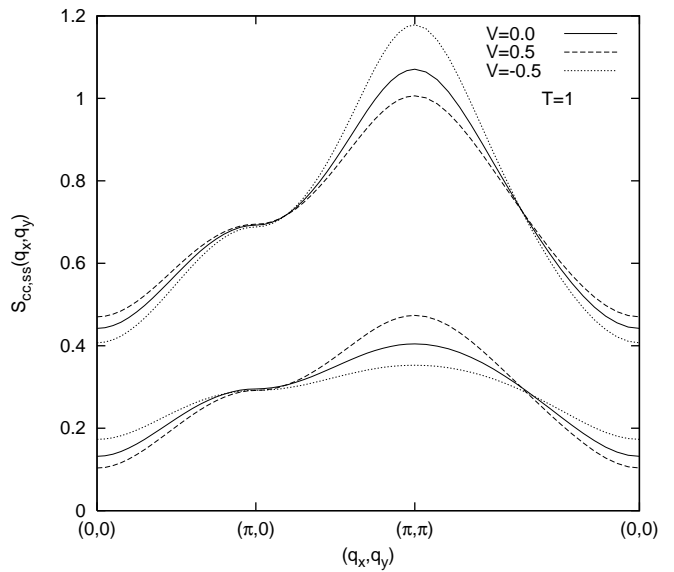}
\end{center}
\caption{The charge and spin components of the structure factors at $T=1$, $%
n=1,\;U=4,$ $V=-0.5,\;0\;$and$\;0.5$. The upper and lower curves correspond
respectively to the spin and charge components.}
\end{figure}

We finally turn to our main point, a more general overview of the effect of
the nearest-neighbor interaction $V$ over a wide range of parameters. In
Fig.~5 we show the variation of $g_{\sigma \tilde{\sigma}}(0)$ and $%
g_{cc,ss}(a)$ as a function of temperature for $V=0,\;0.5\;$and$\;1$. We
first notice that at $V=0$, both $g_{\sigma \tilde{\sigma}}(0)$ and $%
g_{ss}(a)$ have a sharp decrease around $T\approx 0.3$. For $g_{\sigma 
\tilde{\sigma}}(0)$ this means a decrease in the probability for finding two
particles at the same place. In other words, there is an increase in the
size of the local moment. The fact that $g_{ss}(a)=(g_{\sigma \sigma
}(a)-g_{\sigma \tilde{\sigma}}(a))/2$ is negative means that the probability
of finding two electrons at a distance $a$ with opposite spins is larger
than finding them there with the same spin. The decrease of $g_{ss}(a)$ with
temperature indicates a reinforcement of this tendency. These results
reflect the tendency toward antiferromagnetic order (staggered spin order).
Long-range spin-density wave order occurs only at zero temperature, as
required by the Mermin-Wagner theorem in a two dimensional system, but the
decrease in $g_{\sigma \tilde{\sigma}}(0)$ and $g_{ss}(a)$ reflects the
beginning of the renormalized-classical regime where the characteristic spin
fluctuation frequency becomes smaller than temperature and where the
antiferromagnetic correlation length begins to increase exponentially \cite%
{Vilk1,Vilk2}. By contrast, in the charge channel $g_{cc}(a)$ does not show
any strong change in the low temperature limit so there is no tendency to
charge density wave order at these values of $V$ within our present approximation
even at very low temperature. We observe that as we
increase the $V$, the staggered spin fluctuations are depressed since $%
g_{\sigma \tilde{\sigma}}(0)$ and $g_{cc,ss}(a)$ do not decrease sharply.

How spin and charge fluctuations are influenced by $V$ is best illustrated
in Fig.~6 where we show the structure factors at $T=1$, for $V=-0.5,\;0\;$and%
$\;0.5$. Both functions show a peak around $q_{x}=\pi $ and $q_{y}=\pi $, a
sign of the tendency towards staggered ordering. It is obvious from the
figure that antiferromagnetic fluctuations are suppressed with increasing $V$
while the charge fluctuations are enhanced. A negative value of $V$ reverses
the trend. For negative $V$, pairing fluctuations should also become important
but they are not considered here.

\begin{figure}[tbp]
\begin{center}
\includegraphics[scale=0.5]{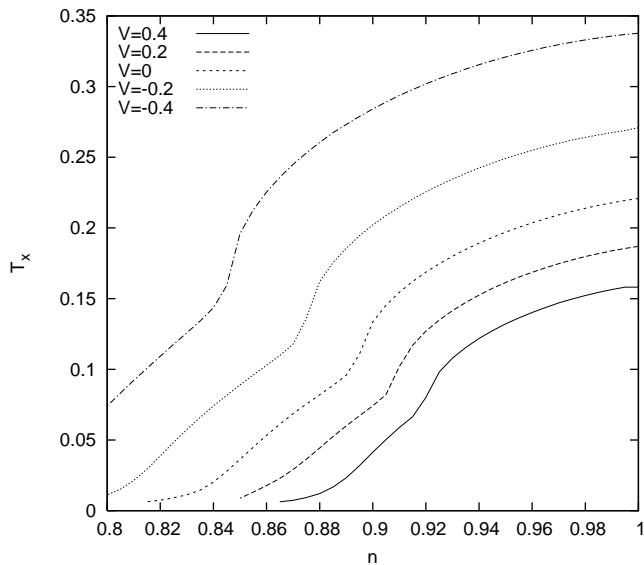}
\end{center}
\caption{The crossover temperature as a function of filling factor for $U=4,$
and $V=-0.4,\;-0.2,\;0,\;0.2,\;0.4$. Positive $V$ reduces the strength of
antiferromagnetic fluctuations while negative $V$ has the opposite trend.}
\end{figure}

\begin{figure}[tbp]
\begin{center}
\includegraphics[scale=0.5]{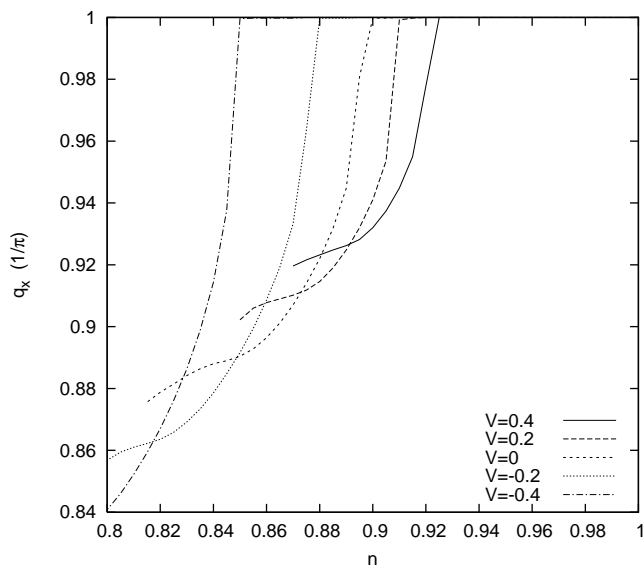}
\end{center}
\caption{The wave-vector where the maximum of the spin response function
appears at the crossover temperature, as a function of filling factor for $%
U=4,$ and $V=-0.4,\;-0.2,\;0,\;0.2,\;0.4$. }
\end{figure}
In Fig.~7 we show the crossover temperature as a function of filling factor
for $V=-0.4,\;-0.2,\;0,\;0.2\;$and$\;0.4$. At this temperature, the
antiferromagnetic correlation length begins to increase exponentially. We
define it as the temperature where the ratio $\chi _{ss}(q_{x},q_{y},0)/\chi
^{0}(q_{x},q_{y},0)$ at its peak reaches the value $100$. The rounding (foot) of
the curves as the crossover temperature vanishes comes from this choice of ratio, 
which corresponds to fixing the correlation length at which we consider that we 
have entered the exponential regime related to the existence of long-range order 
at zero temperature. When the ratio is taken as larger, the rounding occurs over 
a narrower range of densities. In general there are quantum critical points at 
zero temperature where the exponential regime ends\cite{Sachdev}. In Fig.~7, the crossover
temperature and the range of fillings where antiferromagnetic fluctuations
are large both increase or decrease together. It is clear from this figure
that positive $V$ tends to suppress the staggered spin fluctuations while
negative $V$ leads to the opposite trend. As mentioned above, each site has
four nearest-neighbor sites, and our method is quite accurate as long at $%
\left\vert 4V\right\vert <W.$

The apparent breaks in slope on the curves of the last figure correspond to
a change from commensurate to incommensurate fluctuations. We have been
using the terms antiferromagnetic and staggered rather loosely. Fig.~8
corrects this by presenting the peak position at the crossover temperature
of the spin response function (or spin structure factor) as a function of
filling factor for the same parameters as Fig.~7. The maximum in the spin
fluctuations changes with temperature but, at the crossover temperature, it
is at a wave vector that is commensurate near half-filling and then becomes
incommensurate as we decrease $n.$ This comes mainly from the change in the
peak position of the non-interacting susceptibility, but when $V$ is finite
there is an also effect that comes from the presence of the cosine functions
in the effective interaction appearing in the denominator of Eq.~(\ref{chi_ss}%
). The range of fillings where commensurate fluctuations appear is increased
when $V<0$ and decreased when $V>0,$ concomitant with the tendency to
increase or decrease the antiferromagnetic crossover temperature at
half-filling.


\section{Conclusion and Summary}

We generalized the TPSC approach to the extended Hubbard model that contains
nearest-neighbor repulsion $V$ in addition to the usual on-site $U$ term.
That non-perturbative approach, which is a close relative of the STLS
approach used for the electron gas, is valid in the weak to
intermediate-coupling limit. The TPSC approach is usually studied in the
functional derivative formalism and the STLS approach in the Wigner
distribution function formalism, so we presented derivations in both
languages to better illustrate the similarities and differences between the
two approaches.

To derive either TPSC or STLS, two main approximations are necessary: First
we must factor a four-point correlation function (two-body density matrix)
and correct the factorization with the pair correlation function; Second, we
must treat the functional derivative of the pair correlation function with
respect to a fictitious external potential. The STLS factorization of the
pair correlation function does not take into account the range of the
interaction whereas TPSC does. In particular, at $V=0$, the STLS
approximation involves the pair correlation function for all distances
and does not enforce the Pauli principle. On the other hand, TPSC involves
the calculation of $g_{\sigma \widetilde{\sigma }}\left( 0\right)$
only and, in addition, it enforces the Pauli principle $g_{\sigma \sigma }
\left( 0\right) =0.$ The Pauli principle gives an additional equation that
allows an approximate evaluation of the functional derivative entering the
charge channel when one neglects the momentum and frequency dependence of
that functional derivative. The local approximation (where the functional
derivative is replaced by a derivative with respect to density) is another
approach, but it is less accurate than TPSC, as judged from comparisons
with QMC. Analogous comparisons with the spin and charge structure factors
obtained from QMC also show that TPSC is more accurate than STLS.

When $V\neq 0$, one uses the same type of factorization, but extra
functional derivatives appear in TPSC. These extra derivatives cannot be
determined from the same kind of simple sum-rules used for the $V=0$ Hubbard
model. Comparisons of numerical results with QMC simulations show that for $%
4V<W$, one can neglect the extra functional derivatives. In principle, one
could obtain these derivatives within the local approximation or, given an
independent way to obtain the free energy, by enforcing the compressibility
and spin susceptibility sum-rules. This is left for future work.

At $V=0,$ near half filling, there is a crossover temperature to the
renormalized-classical regime where spin fluctuations near $\left( \pi ,\pi
\right) $ grow exponentially, diverging only at zero temperature in
agreement with the Mermin-Wagner theorem in two dimensions. The effect of $%
V>0$ is to decrease both the crossover temperature and the range of dopings
over which this crossover occurs. The crossover to increasing commensurate $%
\left( \pi ,\pi \right) $ fluctuations occurs over a narrower range near $%
n=1 $ when $V>0$ than when $V=0$. The opposite conclusions hold true for $%
V<0 $. All these effects are non-perturbative. They are beyond the reach of
RPA where $V$ does not influence the spin fluctuations. Staggered charge
fluctuations are enhanced by $V>0$ and decreased by $V<0$.


\section{Acknowledgments}

We are indebted to B. Kyung and S. Roy for useful conversations.
This work was supported by NSERC (Canada), FQRNT (Qu\'{e}bec), CFI (Canada),
CIAR and the Tier I Canada Research Chair Program (A.-M.S.T.).


\bigskip

\appendix%

\section{Other possibility for factorization}
\label{App1}

 Here we would like to return to Eq.~(\ref{Self_exact})  in order
to mention the Fock-type factorization of the $V$ term,
which can be introduced as follows
\begin{widetext}
\begin{equation}
\Sigma^{\prime}_\sigma(1,2)=-V\sum_{a}G_\sigma(1,1+a)\delta(1+a,2)g_{xx}(1,1+a)
\label{other_factorization}
\end{equation}
where $\Sigma'_\sigma(1,2)$ is the extra contribution to the self-energy in
Eq.~(\ref{Self_approx}) and $g_{xx}(a)$ is a new pair correlation function
defined as
\begin{equation}
g_{xx}(a)=\frac{\left\langle T_{\tau }c_{\sigma}^{\dagger}(1+a)c_{\sigma }(1+a)
c_{\sigma}^{\dagger }(1)c_{\sigma}(1)\right\rangle }
{G_{\sigma}(1,1+a^+)G_{\sigma}(1+a,1^+)}.%
\label{gxx}
\end{equation}
This pair correlation function is related to the one for parallel spins by
\begin{equation}
g_{xx}(a)=\frac{g_{\sigma\sigma}(a)n_{\sigma}^2}{G_{\sigma}(1,1+a^+)G_{\sigma}(1+a,1^+)}.
\label{gxx-gss}
\end{equation}
The Green function that appeares in the definition of the pair correlation function
Eq.~(\ref{gxx}) can be written as
\begin{align}
G_{\sigma}(1+a,1^+)&=\int\frac{d{\mathbf q}}{\nu}e^{i{\mathbf q}\cdot{\mathbf a}}
n_{\sigma}({\mathbf q})=\int\frac{d{\mathbf q}}{\nu}[\cos({\mathbf q}
\cdot{\mathbf a})+i\sin({\mathbf q}\cdot{\mathbf a})]n_{\sigma}({\mathbf q})
=n_{c}(a)+in_{s}(a).
\label{G1a1}
\end{align}
where $n_{\sigma}({\mathbf q})$ is the momentum distribution $\left\langle
c_{{\mathbf q}\sigma }^{\dagger}c_{{\mathbf q}\sigma}
\right\rangle$.
The extra contribution to the approximate self-energy
is obtained by substituting Eq.~(\ref{G1a1}) and Eq.~(\ref{gxx-gss}) in
Eq.~(\ref{other_factorization}). The final form in Fourier space is
\begin{equation}
\Sigma^{\prime}_{\sigma}({\mathbf q})=-2Vn_{\sigma}^2g_{\sigma\sigma}(a)\frac{n_{c}(a)
[\cos(q_{x}a)+\cos(q_{y}a)]+n_{s}(a)[\sin(q_{x}a)+\sin(q_{y}a)]}{n_{c}(a)^2+n_{s}(a)^2}.
\end{equation}
\end{widetext}%
We stress that the $n_{s}(a)$ term is zero if we assume that the momentum
distribution function (or the dispersion relation) is a symmetric function. In this
case, the final contribution of the above equation renormalizes the hopping
term as follows
\begin{equation}
t^{\prime}=t+\frac{2Vn_{\sigma}^{2}g_{\sigma\sigma}(a)}{n_{c}(a)}.
\end{equation}
The effect of this renormalization  for positive $V$ is merely a depression in
the structure functions. The effect is opposit for negative $V$ and it can even
lead to an instability when $t'\approx 0$. One also should notice that inclusion of $n_{s}(a)$
in the self-energy leads to an asymmetric dispersion relation (or an asymmetric momentum
distribution function) which in return gives a non-zero value for this quantity.
The presence of a self-consistent asymmetry in the momentum distribution function is known as
a Pomeranchuk instability, which can be related to presence of stripes in the system
\cite{Kivelson:1998,Metzner:2003}.

 The extra term in the self-energy Eq.~(\ref{other_factorization}) 
 also leads to an extra
contribution in the response functions. To evaluate this extra correction we
need
\begin{widetext}
\begin{align}
\frac{\delta\Sigma^{\prime}_{\sigma}(4,5)}{\delta G_{\sigma^{\prime\prime}}(6,7)}
=&-V\delta_{\sigma\sigma^{\prime\prime}}
\sum_{a}\delta(4+a,5)\delta(4,6)\delta(4+a,7)g_{xx}(4,4+a)\notag\\
&-V\sum_{a}G_{\sigma}(4,4+a)\delta(4+a,5)
\frac{\delta g_{xx}(4,4+a)}{\delta G_{\sigma^{\prime\prime}}(6,7)}.
\end{align}
\end{widetext}%
By inserting the above equation in Eq.~(\ref{Pi}), after ignoring the second term which involves the
functional derivative, one gets the following extra correction 
to the equation for the response functions
\begin{widetext}
\begin{equation}
\chi_{\sigma\sigma}^{\prime}(1,2)=VG_{\sigma}(1,\bar{3})G_{\sigma}(\bar{3}+a,1)
\frac{\delta G_{\sigma}(\bar{3},\bar{3}+a)}{\delta \phi_{\sigma}(2,2)}
g_{xx}(\bar{3},\bar{3}+a).
\end{equation}
\end{widetext}%
One notices that the three point response function appears in the above
correction for the two body response function. It is hard to
determine this type of function in practice, but its contribution to 
the equation for the two-body response function can be estimated.
Eq.~(\ref{Pi}) for parallel spins is our
basic equation for this task. We replace $1\rightarrow 3$,
$2\rightarrow 3+a$ and $3\rightarrow 1$, then 
multiply both sides of
the equation by $VG_{\sigma}(1,3)G_{\sigma}(3+a,1)g_{xx}(3,3+a)$
and finally performing a sum over the internal index $3$. 
The final result is given by the following equation
\begin{widetext}
\begin{align}
\chi_{\sigma\sigma}^{\prime}(1,2)&=\chi_{\sigma\sigma}^{1V(2)}(1,2)
+Ug_{\sigma\tilde{\sigma}}(0)\chi_{\sigma\sigma}^{1V(2)}(1,\bar{4})\chi_{\tilde{\sigma}\sigma}(\bar{4},2)
\notag\\&+V\sum_{a,\sigma^{\prime}}g_{\sigma\sigma^{\prime}}(a)
\chi_{\sigma\sigma}^{1V(2)}(1,\bar{4}+a)\chi_{\sigma^{\prime}\sigma}(\bar{4}+a,2)\notag\\
&+Un_{\tilde{\sigma}}\sum_{\sigma^{\prime}}\frac{\delta g_{\sigma\tilde{\sigma}}(0)}
{\delta n_{\sigma^{\prime}}}\chi_{\sigma\sigma}^{1V(2)}(1,\bar{4})\chi_{\sigma^{\prime}\sigma}(\bar{4},2)
\label{chiprime}
\end{align}
where $\chi_{\sigma\sigma}^{1V(2)}(1,2)$ is
\begin{equation}
\chi_{\sigma\sigma}^{1V(2)}(\mathbf{q},\omega_{n})=Vg_{xx}(a)G_{\sigma}(1,\bar{3})
G_{\sigma}(\bar{3},2)G_{\sigma}(1,\bar{3}+a)G_{\sigma}(\bar{3}+a,2)
\end{equation}
\end{widetext}%
while $1V$ stands for the first order in $V$ and $(2)$ is a label for its equivalent diagram.
One can replace $\chi_{\sigma\sigma^{\prime}}(1,2)$ in Eq.~(\ref{chiprime}) which
already contains $\chi_{\sigma\sigma}^{\prime}(1,2)$. Eq.~({\ref{chiprime}})
and Eq.~({\ref{Piccss}}) after equating labels $1$ and $2$ 
[in Eq.~({\ref{Piccss}})] form a set of coupled
equations which in principle can be solved. Here we are not going to provide the final form of
the equations and instead simply explain that the correction term is small compared to its counterparts.
This can be easily understood from Eq.~(\ref{chiprime}) because $\chi_{\sigma\sigma}^{1V(2)}(1,2)$
appears as a multiplicative factor and also as a separate factor in all terms, so having
an estimate from this term tells us a lot about the importance of the correction
term. This term can be easily evaluated and it has the following form in Fourier space
\begin{widetext}
\begin{equation}
\chi_{\sigma\sigma}^{1V(2)}(\mathbf{q},\omega_{n})=g_{xx}(a)
\int\frac{d\mathbf{k}}{\nu}\int\frac{d\mathbf{k}^{\prime}}{\nu}
\frac{f_{\sigma}(\mathbf{k}+\frac{\mathbf{q}}{2})-f_{\sigma}(\mathbf{k}-\frac{\mathbf{q}}{2})}{i\omega_{n}
-(\epsilon_{\mathbf{k}+\frac{\mathbf{q}}{2}}-\epsilon_{\mathbf{k}-\frac{\mathbf{q}}{2}})}
\frac{f_{\sigma}(\mathbf{k}^{\prime}+\frac{\mathbf{q}}{2})-f_{\sigma}(\mathbf{k}^{\prime}-\frac{\mathbf{q}}{2})}
{i\omega_{n}-(\epsilon_{\mathbf{k}^{\prime}+\frac{\mathbf{q}}{2}}-\epsilon_{\mathbf{k}^{\prime}-\frac{\mathbf{q}}{2}})}
V(|\mathbf{k}-\mathbf{k}^{\prime}|),
\label{X1V(2)}
\end{equation}
\end{widetext}%
where $V(q)=V[\cos(q_{x}a)+\cos(q_{y}a)]$. The above function is proportional 
to one of the first order diagrams that appear in the perturbation expansion of the response functions. 
We evaluated the above equation numerically and compared the result with the other first order diagram, 
which is given by
\begin{equation}
\chi_{\sigma\sigma^{\prime}}^{1V(1)}(\mathbf{q},\omega_{n})=g_{\sigma\sigma^{\prime}}(a)V(q)
\chi^{0}_{\sigma\sigma}(\mathbf{q},\omega_{n})^{2}.
\label{X1V(1)}
\end{equation}
We checked that Eq.~(\ref{X1V(2)}) is zero at half-filling within our numerical precision and
its contribution away from half-filling is negligible compared to Eq.~(\ref{X1V(1)}).

\section{Improved self-energy (second step of the TPSC approximation)}

We can improve our approximation for the self-energy to include
single-particle scattering off low-energy spin and charge fluctuations.
These processes give momentum and frequency dependence to the self-energy
\cite{Vilk3, Moukouri:2000, Allen:2003, Vilk2}. This improved self-energy
leads to one-particle spectral functions that compare extremely well with
QMC in the case of the usual Hubbard model \cite{Moukouri:2000}. In the
extended Hubbard model, the improved self-energy that includes the effects
of longitudinal fluctuations can be obtained as follows. We first write the
exact result
\begin{widetext}%
\begin{align}
\Sigma _{\sigma }(1,\bar{2})G_{\sigma }(\bar{2},3)=& -U\left\langle T_{\tau
}c_{\tilde{\sigma}}^{\dagger }(1)c_{\tilde{\sigma}}(1)c_{\sigma
}(1)c_{\sigma }^{\dagger }(3)\right\rangle  \notag \\
& -V\sum_{\sigma ^{\prime },a}\left\langle T_{\tau }c_{\sigma ^{\prime
}}^{\dagger }(1+a)c_{\sigma ^{\prime }}(1+a)c_{\sigma }(1)c_{\sigma
}^{\dagger }(3)\right\rangle  
\label{Sigma1} \\
=& -U\left[ \frac{\delta G_{\sigma }(1,3)}{\delta \phi _{\tilde{\sigma}%
}(1^{++},1^{+})}-G_{\tilde{\sigma}}(1,1^{+})G_{\sigma }(1,3)\right]  \notag
\\
& -V\sum_{\sigma ^{\prime },a}\left[ \frac{\delta G_{\sigma }(1,3)}{\delta
\phi _{\sigma ^{\prime }}(1+a^{++},1+a^{+})}-G_{\sigma ^{\prime
}}(1+a,1+a^{+})G_{\sigma }(1,3)\right].
\label{Sigma}
\end{align}%
After replacing $3\rightarrow 1^{+}$ and using rotational symmetry, we have
\begin{align}
\Sigma_{\sigma}(1,\bar{2}) G_{\sigma}(\bar{2},1^{+})
&=Un_{\sigma}n_{\tilde{\sigma}}g_{\sigma\tilde{\sigma}}(0)+2Vn^{2}g_{cc}(a) \label{SROPL1}\\
&=U[\chi_{\sigma\tilde{\sigma}}(1,1)+n_{\sigma}n_{\tilde{\sigma}}]
+2V[\chi_{cc}(1,1+a)+n^{2}]
\label{SROPL}
\end{align}
\end{widetext}%
which can be interpreted as a sum-rule relating one particle
quantities to the left and two-particle quantities to the right.
Note that the equalities
\begin{align}
g_{\sigma\tilde{\sigma}}(0)&=\frac{\chi_{\sigma\tilde{\sigma}}(1,1)}
{n_{\sigma}n_{\tilde{\sigma}}}+1,\notag\\
g_{cc}(a)&=\frac{\chi_{cc}(1,1+a)}{n^{2}}+1
\label{FDTRS}
\end{align}
hold. They are the real-space version of the
fluctuation-dissipation theorem.

We already approximated the self-energy using a factorization of Eq.~(\ref{Sigma1}).
Our factorization is exact in the limit $3\rightarrow 1^{+}$ so
we don't expect anything new from that equation. However, we can still use an approximate
form for the response functions in the second identity Eq.~(\ref{Sigma}). Now our main point is that if
the response functions satisfy the sum-rules at the two-particle level Eq.~(\ref{FDTRS}),
then there is a guarantee that the self-energy satisfies the sum-rule in
Eq.~(\ref{SROPL1}) relating one- and two-particle quantities.

 As an example one can insert Eq.~(\ref{Pi}) inside Eq.~(\ref{Sigma}), which already contains
the approximate form of the self-energy Eq.~(\ref{Self_approx}), to get an approximate form
of the self energy at a second level of approximation.
We give the final form of this equation in
Fourier space for practical use:
\begin{widetext}%
\begin{align}
\Sigma _{\sigma }(\mathbf{k},\omega _{n})\approx  (Un_{\tilde{\sigma}}+4Vn)
&+\frac{T}{4}\sum_{\omega _{n^{\prime}}}\int\frac{d\mathbf{q}}{\nu}\{UU_{ss}(\mathbf{q})\chi _{ss}(\mathbf{%
q},\omega _{n^{\prime }})\notag\\&+U_{cc}(\mathbf{q})[U+4V\gamma(\mathbf{q})]\chi _{cc}(\mathbf{q},
\omega _{n^{\prime }})\}G_{0}(\mathbf{k+q},\omega _{n}+\omega_{n^{\prime }}).
\label{Sigmak}
\end{align}%
\end{widetext}%
where $\gamma(\mathbf{q})=\sum_{\alpha}\cos(q_{\alpha}a)$.
As we expect, the above formula for the self-energy Eq.~(\ref{Sigmak}) satisfies the
sum-rule in Eq.~(\ref{Sigma1}). This, in fact, is a result of using
Eqs.~(\ref{gcc}), (\ref{gss}), (\ref{FT}) and (\ref{FDT}), which are
another version of the fluctuation-dissipation theorem Eq.~(\ref{FDTRS}).

We have to mention that we dropped out a term during the
calculation which is given by
\begin{equation}
\Sigma^{\prime}_{\sigma HF}(1,2)=-V\sum_{a}G_\sigma(1,1+a)\delta(1+a,2).
\end{equation}
One should include the correction term to the response function we already discussed in the
last section to take care of presence the above term. This means that one should include both
corrections in the self-energy and the response function (or ignore them from both) to have the
sum-rule Eq.~(\ref{SROPL1}).

Following the procedure established for the ordinary Hubbard
model \cite {Moukouri:2000, Allen:2003}, one could also take into account transverse
spin fluctuations and crossing symmetry to write a more general result.


\newpage



\begin{thebibliography}{99}
\bibitem{Vilk1} Y. M. Vilk, Liang Chen and A.-M.S. Tremblay, Phys. Rev. B
\textbf{49}, 13267 (1994); Y.M. Vilk, Liang Chen, and A.-M.S. Tremblay,
Physica C \textbf{235-240}, 2235 (1994).

\bibitem{Vilk2} Y. M. Vilk, and A. -M.S. Tremblay, J. Phys. I \textbf{7},
1309 (1997).

\bibitem{Moukouri:2000} S. Moukouri, S. Allen, F. Lemay, B. Kyung, D.
Poulin, Y.M. Vilk et A.-M. S. Tremblay, Phys. Rev. B \textbf{61}, 7887
(2000).

\bibitem{Kyung:2000} B. Kyung, S. Allen, A.-M. S. Tremblay, Phys. Rev. B
\textbf{64}, 075116 (2001).

\bibitem{Kyung:2003} B. Kyung, J.S. Landry, D. Poulin, A.-M.S. Tremblay,
Phys. Rev. Lett. \textbf{90}, 099702 (2003).

\bibitem{Kyung:2003b} B. Kyung, J.S. Landry and A.-M.S. Tremblay, Phys. Rev.
B \textbf{68}, 174502 (2003).\qquad

\bibitem{Kanamori-Bruckner} J. Kanamori, Prog. Theor. Phys. \textbf{30}, 275 
(1963); K. A. Bruckner and C. A. Levinson, Phuys. Rev. \textbf{97}, 2344 (1955);
K. A. Bruckner and J. L. Gummel, Phys. Rev. \textbf{109}, 1023 (1958); {\it ibid}
\textbf{109}, 1040 (1958).

\bibitem{Singwi} K. S. Singwi, M. P. Tosi, R. H. Land and A. Sj\"{o}lander,
Phy. Rev. \textbf{176}, 589 (1968).

\bibitem{fp} Some of these higher-order correlation functions will be discussed 
more deeply in a future publication.

\bibitem{Imada:2005} Kota Hanasaki, and Masatoshi Imada, cond-mat/0506240.

\bibitem{Onari:2004} S. Onari, R. Arita, K.Kuroki and H.Aoki, Phys. Rev. B 
\textbf{70, }094523 (2004).

\bibitem{Pietig:1999} R. Pietig, R. Bulla, and S. Blawid. Phys. Rev. Lett. 
\textbf{82, }4046 (1999).

\bibitem{Calandra:2002} M. Calandra, J. Merino, and R. H. Mckenzie, Phys.
Rev. B. \textbf{66} 195102 (2002).

\bibitem{Vojta:1999} M. Vojta, R.E. Hetzel, R.M. Noack, Phys. Rev. B \textbf{%
60}, R8417 (1999).

\bibitem{Vojta:2001} Vojta M, Hubsch A, Noack RM, Phys. Rev. B \textbf{63},
045105 (2001).

\bibitem{Fourcade} B. Fourcade and G. Sproken, Phys. Rev. B \textbf{29},
5096 (1984);

\bibitem{Bosch:1988} L. M. del Bosch and L. M. Falicov, Phys. Rev. B \textbf{%
37}, 6037 (1988).

\bibitem{Aichhorn:2004} M. Aichhorn, H.G. Evertz, W. von der Linden, and M.
Potthoff, Phys. Rev. B \textbf{70,} 235107 (2004).

\bibitem{Avella:2004} A. Avella, F. Mancini, Euro. Phys. J. B \textbf{41},
149 (2004).

\bibitem{Ohta:1994} Y. Ohta, K. Tsutsui, W. Koshibae, and and S. Maekawa,
Phys. Rev. B \textbf{50}, 13594 (1994).

\bibitem{Yan:1993} X.Z. Yan Phys. Rev. B \textbf{48}, 7140 (1993).

\bibitem{Barktowiak:1995} B M. Bartkowiak, J. A. Henderson, J. Oitmaa, P. E.
de Brito, Phys. Rev. B \textbf{51}, 14077 (1995).

\bibitem{VanDongen:1994} P. G. J. van Dongen, Phys. Rev. B \textbf{49}, 7904
(1994); P. G. J. van Dongen, Phys. Rev. B \textbf{50}, 14016 (1994).

\bibitem{Onari} S. Onari, R. Arita, K. Kuroki, and H. Aoki, Phys. Rev. B
\textbf{70}, 094523 (2004).

\bibitem{Kishine:1995} J. Kishine, H. Namaizawa, Prog. Theor. Phys. \textbf{%
93}, 519 (1995).

\bibitem{Callaway:1990} J. Callaway, D.P. Chen, D.G. Kanhere, and Qiming Li,
Phys. Rev. B \textbf{42}, 465 (1990) and Physica B \textbf{163}, 127 (1990).

\bibitem{Sachdev} Subir Sachdev, {\it Quantum Phase Transition} (Camdridge University Press, 1999).

\bibitem{Kivelson:1998} S. A. Kivelson, E. Fradkin, and V. J. Emery, Nature 
(London) \textbf{393}, 550 (1998).

\bibitem{Metzner:2003} A. Neumayer and  W. Metzner, Phys. Rev. B \textbf{67},
035112.

\bibitem{Zhang} Y. Zhang and J. Callaway, Phys. Rev. B \textbf{39}, 9397
(1989).

\bibitem{Allen:2003} S. Allen, A.-M.S. Tremblay, Y.M. Vilk, in "Theoretical
Methods for Strongly Correlated Electrons", David S\'{e}n\'{e}chal, Andr\'{e}%
-Marie Tremblay and Claude Bourbonnais (eds.) CRM Series in Mathematical
Physics, (Springer, New York, 2003), p.341.

\bibitem{Baym:1962} Gordon Baym, Phys. Rev. \textbf{127}, 1391 (1962).

\bibitem{BaymKadanoff:1962} L. P. Kadanoff and G. Baym, \textit{Quantum
Statistical Mechanics} (Benjamin, Menlo Park, 1962).

\bibitem{MartinSchwinger:1959} P.C. Martin and J. Schwinger, Phys. Rev.
\textbf{115}, 1342 (1959). This paper also contains numerous references to
previous work.

\bibitem{Vilk3} Y.M. Vilk and A.-M.S. Tremblay, J. Phys. Chem. Solids
\textbf{56}, 1769 (1995); Y.M. Vilk, and A.-M.S. Tremblay, Europhys. Lett.
\textbf{33}, 159 (1996).

\bibitem{Vashishta} P. Vashishta and K. S. Singwi, Phys. Rev. B \textbf{6} ,
875 (1972).

\bibitem{Moudgil} R. K. Moudgil, P. K. Ahluwalia and K. N. Pathak, Phys.
Rev. B \textbf{52}, 11945 (1995).

\bibitem{Nelisse:1999} In one dimension, the momentum dependence of the
interactions cannot be neglected. See for example, H. N\'{e}lisse, C.
Bourbonnais, H. Touchette, Y.M. Vilk and A.-M.S. Tremblay. European Journal
of Physics B \textbf{12}, 351 (1999).

\bibitem{Roy:2005} S. Roy, C. Brillon and A.-M.S. Tremblay (unpublished).

\bibitem{Hirsch} J. E. Hirsch, Phys Rev Lett. \textbf{53}, 2327 (1984); H.
Q. Lin and J. E. Hirsch, Phys. Rev. B \textbf{33}, 8155, (1986).
\end{thebibliography}
\end{document}